\documentclass[fleqn,10pt]{wlscirep}
\usepackage[utf8]{inputenc}
\usepackage[T1]{fontenc}

\usepackage{xcolor}

\title{Unraveling the Geometry of Visual Relational Reasoning}

\author[1]{Jiaqi Shang}
\author[2,3]{Gabriel Kreiman}
\author[3,4,*]{Haim Sompolinsky}
\affil[1]{Program in Neuroscience, Harvard Medical School, Boston, 02115, Massachusetts, United States}
\affil[2]{Boston Children’s Hospital, Harvard Medical School, Boston, 02115, Massachusetts, United States}
\affil[3]{Center for Brains, Minds, and Machines, Cambridge, 02139, Massachusetts, United States}
\affil[4]{Edmond and Lily Safra Center for Brain Sciences, Hebrew University, Jerusalem, 9190401, Israel}

\affil[*]{hsompolinsky@mcb.harvard.edu}

\begin{abstract}
Humans readily generalize abstract relations, such as recognizing “constant” in shape or color, whereas neural networks struggle, limiting their flexible reasoning. To investigate mechanisms underlying such generalization, we introduce \mbox{\textit{SimplifiedRPM}}, a novel benchmark for systematically evaluating abstract relational reasoning, addressing limitations in prior datasets. In parallel, we conduct human experiments to quantify relational difficulty, enabling direct model-human comparisons. Testing four models—ResNet-50, Vision Transformer, Wild Relation Network, and Scattering Compositional Learner (SCL)—we find that SCL generalizes best and most closely aligns with human behavior. Using a geometric approach, we identify key representation properties that accurately predict generalization and uncover a fundamental trade-off between signal and dimensionality: novel relations compress into training-induced subspaces. Layer-wise analysis reveals where relational structure emerges, highlights bottlenecks, and generates concrete hypotheses about abstract reasoning in the brain. Motivated by these insights, we propose SNRloss, a novel objective explicitly balancing representation geometry. Our results establish a geometric foundation for relational reasoning, paving the way for more human-like visual reasoning in AI and opening promising avenues for extending geometric analysis to broader cognitive tasks.
\end{abstract}
\begin{document}

\flushbottom
\maketitle
%
%
\thispagestyle{empty}

\section*{Introduction}

Humans exhibit a remarkable ability to recognize and generalize abstract relations across diverse contexts. Once we understand an abstract relation like “constant,” we can flexibly apply it to different object attributes,  such as a row of items sharing the same color, shape, or number. This generalization ability allows us to recognize patterns and make inferences in novel situations, supporting flexible reasoning and complex problem-solving.

A central challenge in both cognitive science and artificial intelligence is to understand the mechanisms that enable this kind of generalization. Raven’s Progressive Matrices (RPM) task~\cite{raven1936mental} has long served as a benchmark for abstract relational reasoning. In this task, participants view a $3 \times 3$ matrix of panels with the bottom-right panel missing. Participants must infer the relational rules governing the rows or columns, such as “constant shape across the row”, and use these inferred rules to select the missing panel from several alternatives. Decades of behavioral studies in cognitive science revealed consistent patterns of human performance on the task and have established RPM accuracy as a reliable predictor of general cognitive and problem-solving abilities
~\cite{raven1936mental,forbes1964item,carpenter1990one,matzen1994error,matzen2010recreating}. Yet, the cognitive mechanisms that support this generalization ability remain unclear.

Inspired by RPM, artificial intelligence research has developed large-scale tasks, such as PGM~\cite{barrett2018measuring} and RAVEN~\cite{zhang2019raven, hu2021stratified}, to evaluate relational reasoning in neural networks. While models trained on these datasets~\cite{barrett2018measuring, zhang2019raven, hu2021stratified, wu2020scattering, webb2020emergent, webb2024relational, steenbrugge2018improving, hersche2023neuro, rahaman2021dynamic, zhang2022learning, zhang2019learning, zhang2024human, mondal2024slot} achieve high accuracy on learned relational rules, they generalize poorly to novel rules~\cite{malkinski2022deep}. Recent evaluations of large-scale pre-trained vision-language models on the RAVEN task have shown that these models also struggle, despite employing advanced inference strategies such as in-context learning~\cite{zhang2024far}. This limitation is further compounded by the lack of detailed human behavioral data on the same tasks, making it challenging to assess whether the model's behavior aligns with human reasoning. While recent work has begun to explore mechanisms in simplified relational tasks, such as transitive inference~\cite{miconi2025neural, lippl2024mathematical, kay2024emergent} and same-different relations~\cite{fleuret2011comparing, zerroug2022benchmark, kim2018not}, broad generalization across diverse relations and realistic image inputs remains an open problem.

This gap motivates a fundamental question: What mechanisms enable or constrain the generalization of abstract relations? Specifically, how are relational rules represented in neural networks, and what geometric principles govern whether they can be flexibly applied to new contexts?

To address these questions, we introduce \textit{SimplifiedRPM}, a novel benchmark designed to systematically evaluate abstract relational generalization. We assess both human performance and a range of neural network models on this benchmark. We propose that a model’s ability to generalize can be understood in terms of the geometry of its emergent rule manifolds—the representations formed for different relational rules. 

To formalize this hypothesis, we extend a prior geometric framework developed for object classification~\cite{sorscher2022neural} to the domain of relational reasoning. We show that generalization performance is accurately predicted by several measurable and interpretable rule manifold geometries. Our analysis reveals a core tradeoff: generalization is constrained by a compression mechanism that restricts both seen and unseen rule manifolds to a shared low-dimensional subspace shaped by the training rules. Motivated by these insights, we propose SNRloss, a novel training objective that directly shapes representational geometry to improve relational generalization.

Together, our findings provide a principled geometric account of abstract relation generalization, offer testable hypotheses for abstract relation reasoning in the brain, and pave the way for applying geometric analysis to broader domains in cognitive science and artificial intelligence.

\section*{Results}

\subsection*{Evaluating Deep Neural Networks in Relational Reasoning}
\begin{figure}[ht]
\centering
\includegraphics[width=\linewidth]{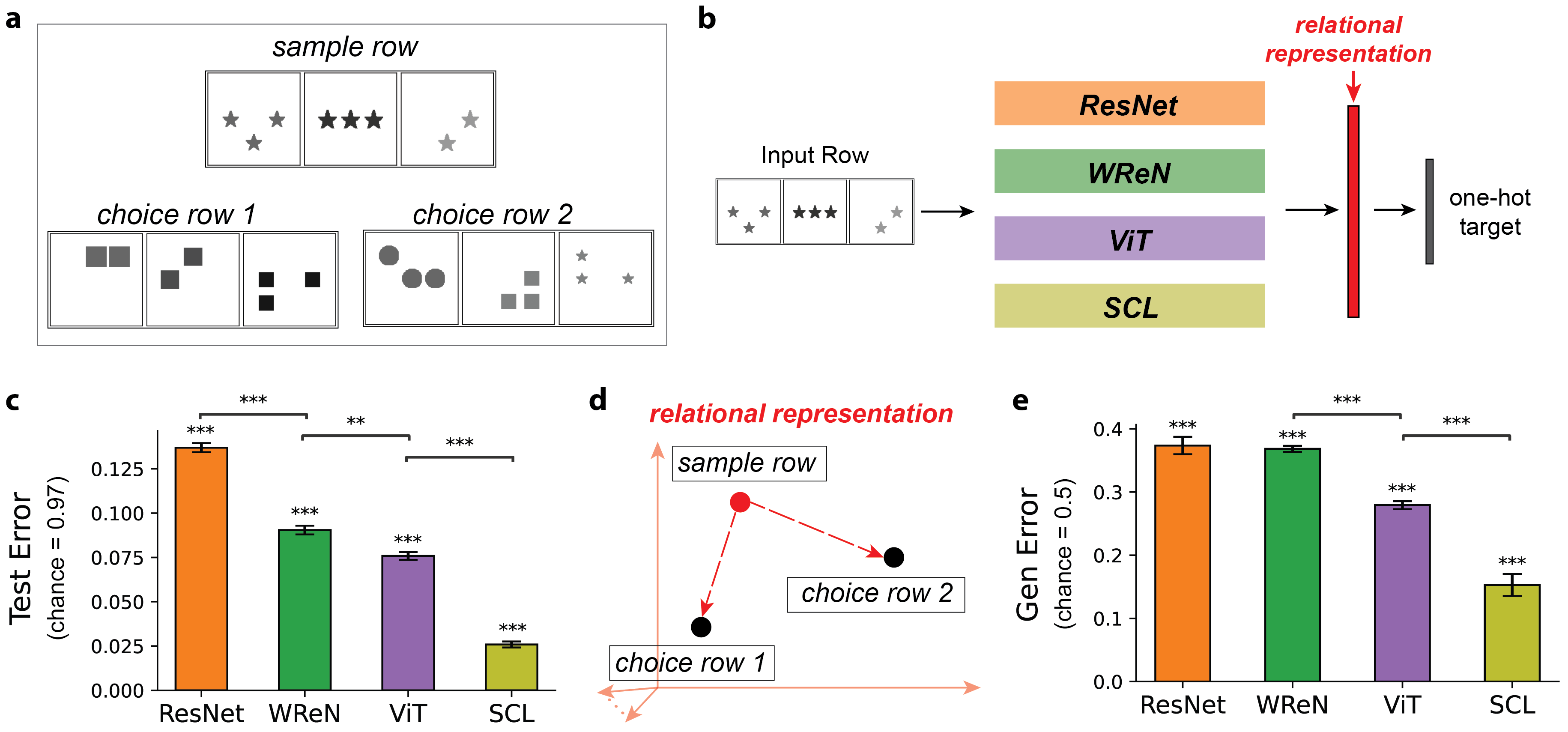}
\caption{The \textit{SimplifiedRPM} dataset helps evaluate neural networks in relational reasoning and generalization. 
        (a) Example \textit{SimplifiedRPM} trial with ``constant shape`` rule. Choice 1 is correct; Choice 2 follows ``constant number'' and is incorrect.
        (b) We evaluate four representative models on the \textit{SimplifiedRPM} task: ResNet-50 (orange)~\cite{he2016deep}, WReN (green)~\cite{barrett2018measuring}, ViT (purple)~\cite{dosovitskiy2020image}, and SCL (yellow-green)~\cite{wu2020scattering}. Each model encodes relational rules in rows using high-dimensional relational representations (red) and is trained end-to-end. A linear classifier is appended during training to categorize relational rules. See Methods for details.
        (c) Test errors for four models across 15 random splits of training and held-out rules, with error bars denoting s.e.m. All models perform significantly above chance ($0.97$) based on a one-sample t-test: ResNet ($t(14)=-312.77$, $p=1.2\times10^{-28}$), WReN ($t(14)=-343.78$, $p=3.4\times10^{-29}$), ViT ($t(14)=-388.11$, $p=6.2\times10^{-30}$) and SCL ($t(14)=-542.53$, $p=5.7\times10^{-32}$). Pairwise Wilcoxon tests confirm SCL achieves the lowest test error, followed by ViT, WReN, and ResNet ($p$-values: SCL vs. ViT, $3.0\times10^{-5}$; ViT vs. WReN, $1.6\times10^{-3}$; WReN vs. ResNet, $3.0\times10^{-5}$). 
        (d) Schematic illustrating distances (red arrows) between sample and choice row relational representations. The trial is solved correctly if the correct choice row representation is closer to the sample row than the incorrect choice.
        (e) Generalization error of four models on \textit{SimplifiedRPM}, averaged over $15$ random training and held-out rule splits. Error bars denote s.e.m. All models perform significantly above chance ($0.5$) based on a one-sample t-test: ResNet ($t(14)=-8.93$, $p=1.8\times10^{-7}$), WReN ($t(14)=-26.39$, $p=1.2\times10^{-13}$), ViT ($t(14)=-33.45$, $p=4.6\times10^{-15}$), and SCL ($t(14)=-19.35$, $p=8.3\times10^{-12}$). Pairwise Wilcoxon tests confirm SCL achieves the lowest error, followed by ViT, while WReN and ResNet show similarly high errors ($p$-values: SCL vs. ViT, $6.1\times10^{-5}$; ViT vs. WReN, $3.0\times10^{-5}$). Asterisks indicate statistical significance: ** ($p < 0.01$), *** ($p < 0.001$).}
\label{fig:Fig1_RPMtask_and_perf}
\end{figure}

To systematically assess neural networks’ ability to generalize abstract relations, we developed the Simplified Raven’s Progressive Matrices (\textit{SimplifiedRPM}) task (Fig.~\ref{fig:Fig1_RPMtask_and_perf}a). Inspired by the original RPM task~\cite{raven1936mental}, \textit{SimplifiedRPM} requires models to infer a relational rule from a \textit{sample row} (Fig.~\ref{fig:Fig1_RPMtask_and_perf}a, top) and determine which of two \textit{choice rows} (Fig.~\ref{fig:Fig1_RPMtask_and_perf}a, bottom) follows the same rule. Each row consists of three \textit{panels} with abstract objects, such as triangles or circles, arranged according to a relational rule. For example, the sample row in Fig.~\ref{fig:Fig1_RPMtask_and_perf}a illustrates the ``constant shape'' rule, where all objects share the same shape while other attributes, such as number or position, vary. \textit{SimplifiedRPM} refines the original RPM task~\cite{raven1936mental} by reducing answer choices from eight to two and presenting a single example row instead of two to remove multi-row comparisons. These changes maintain RPM's core challenge while enabling a more focused assessment of relational reasoning.

We construct a dataset for evaluating abstract relation generalization in the \textit{SimplifiedRPM} task, comprising rows governed by 40 distinct relational rules. Each rule applies an abstract relation (e.g., constant) to an object attribute (e.g., shape). The abstract relations and attributes are selected to be consistent with those used in human experiments~\cite{raven1936mental, matzen1994error} while maximizing the number of rules for comprehensive evaluation. Importantly, each row corresponds to a unique rule. We generate many rows per rule, resulting in a dataset of 400k training rows, 40k validation rows, and 40k test rows. To construct a \textit{SimplifiedRPM} task trial, a sample row is selected that follows one rule. A correct choice is drawn from the same rule, while an incorrect choice is drawn from a different rule. Further details are provided in the Methods section.

We assess models' ability to generalize abstract relations using a held-out rule split, in which $5$ out of $40$ relational rules are excluded from training. During training, models learn relational structures from the remaining $35$ rules. For instance, a model trained on "constant number" and "OR shape" may never encounter "constant shape." After training, their performance is evaluated on the five withheld rules using the \textit{SimplifiedRPM} task. Instead of evaluating models pertained on large-scale datasets, where the overlap between their training distribution and the tested rules is ambiguous, we train models under this controlled rule-split setup. This approach ensures precise control over the relational rules encountered during training, enabling a rigorous test of relational generalization. This addresses limitations in existing RPM benchmarks: RAVEN~\cite{zhang2019raven} does not include splits where test rules are entirely unseen during training, and although PGM's Held-out Triplet split withholds some rules, test items still combine seen and unseen rules, allowing models to rely on familiar patterns. By systematically varying which rules are withheld, our benchmark provides a comprehensive assessment of models’ ability to generalize across different types of abstract relations.

We evaluate four representative models on the \textit{SimplifiedRPM} task: ResNet-50~\cite{he2016deep} and ViT~\cite{dosovitskiy2020image}, two benchmark vision models, and WReN~\cite{barrett2018measuring} and SCL~\cite{wu2020scattering}, which are explicitly designed for relational reasoning. These models employ different computational strategies: ResNet-50 applies a series of convolutional layers across all panels simultaneously, WReN computes pairwise panel relationships and aggregates them to infer underlying rules, ViT leverages self-attention to capture relational dependencies, and SCL employs a shared relational module to improve abstract relation recognition across object attributes. Together, these models represent diverse approaches to solving RPM tasks. All models are randomly initialized and trained end-to-end. Further details on the model architecture and hyperparameters are provided in the Methods section.

The models process a row to generate a high-dimensional \textit{relational representation} of its underlying relational rule (Fig.~\ref{fig:Fig1_RPMtask_and_perf}b). To train the models on $35$ relational rules, we appended a linear classification layer to the relational representations and optimized them end-to-end using cross-entropy loss with one-hot rule labels. Fig.~\ref{fig:Fig1_RPMtask_and_perf}c shows test errors across $15$ random training and held-out rules splits. All models successfully learned the relational rules, achieving test errors well below the chance level of $34/35=0.97$. Among them, SCL achieved the lowest test error, followed by ViT, WReN, and ResNet-50.

We assess the trained models’ ability to generalize abstract relations to unseen rules using the \textit{SimplifiedRPM} task. While human reasoning in RPM tasks likely involves complex cognitive processes, we follow prior work in using classification to quantify relational generalization in neural networks. This serves as a useful proxy, linking internal representations to task performance. For each pair in the five held-out rules, we generate $500$ trials by randomly selecting a sample row that follows one rule, pairing it with a correct choice from the same rule and an incorrect choice from a different rule. In each trial, the model is sequentially presented with the sample row, followed by the two choice rows. The model computes the Euclidean distance between the relational representation of the sample row and each choice row. A trial is deemed correct if the correct choice row is closer to the sample row than the incorrect one (Fig.~\ref{fig:Fig1_RPMtask_and_perf}d). All models outperformed the chance generalization error of $0.5$ (Fig.~\ref{fig:Fig1_RPMtask_and_perf}e). Among them, SCL achieved the lowest generalization error, followed by ViT, WReN, and ResNet. Notably, SCL also outperformed the other models on the PGM benchmark’s held-out rule split~\cite{malkinski2022deep}, aligning with our observed generalization order.

\subsection*{SCL Captures Human-like Differential Relational Rule Difficulties}

\begin{figure}[ht]
\centering
\includegraphics[width=\linewidth]{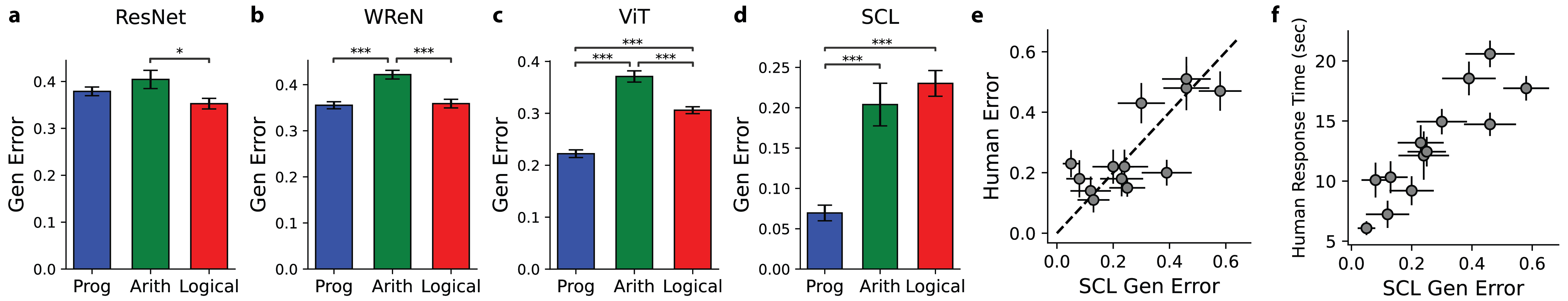}
\caption{SCL Captures Human-like Differential Relational Rule Difficulties.
        (a-d) Generalization errors of four models on the \textit{SimplifiedRPM} task, evaluated across three abstract relation types: Progression (blue), Arithmetic (green), and Logical (red). The chance level is 0.5. Each bar represents the average error for trials where the correct rule of the sample row fell into one of these relation types, with error bars denoting s.e.m.
        (a) ResNet exhibited significantly higher errors for arithmetic rules than logical rules (p = 0.019, Pairwise Wilcoxon test).
        (b) WReN showed significantly higher errors for arithmetic rules than both progression ($p=4.2 \times 10^{-6}$) and logical ($p=9.1 \times 10^{-5}$) rules.
        (c) ViT exhibited higher errors for arithmetic rules compared to both progression ($p=3.8 \times 10^{-17}$) and logical ($p=7.7 \times 10^{-7}$) rules. Additionally, logical rules had significantly higher errors than progression ($p=2.0 \times 10^{-12}$).
        (d) SCL progression was lower than arithmetic ($p=2.1 \times 10^{-17}$) and logical ($p=3.2 \times 10^{-8}$). Statistical significance was assessed using Pairwise Wilcoxon tests. Asterisks indicate statistical significance: * ($p < 0.01$), *** ($p < 0.001$).
        (e) Comparison of the SCL model's generalization errors against human performance on the \textit{SimplifiedRPM} task across 13 selected rule pairs (chance level = 0.5). Each dot represents the average model (x-axis) or human (y-axis) error over 10 questions for a given rule pair, with error bars indicating the s.e.m. The dashed line represents the line of identity.
        (f) Correlation between the SCL model's generalization error (x-axis) and human response times (y-axis) for the 13 selected rule pairs. Each dot represents the average model error and corresponding human response time for a given rule pair, with error bars indicating the s.e.m. }
\label{fig:Fig2_perf_per_rule}
\end{figure}

Analyzing only the averaged generalization error is insufficient to fully assess neural networks' ability to generalize abstract relations, as a model may perform well overall but struggle with rules of specific relation types. To gain deeper insight, we analyzed generalization errors separately for trials where the correct rule of the sample row fell into one of three abstract relation types: progression, arithmetic, or logical. We observed distinct error patterns across models (Fig.~\ref{fig:Fig2_perf_per_rule}a-d). ResNet had higher errors for arithmetic than logical rules (Fig.~\ref{fig:Fig2_perf_per_rule}a). WReN showed the highest errors for arithmetic, performing worse than progression and logical rules (Fig.~\ref{fig:Fig2_perf_per_rule}b). ViT struggled most with arithmetic, while logical rules had higher errors than progression (Fig.~\ref{fig:Fig2_perf_per_rule}c). In contrast, SCL showed similarly high errors for arithmetic and logical rules but significantly lower errors for progression (Fig.~\ref{fig:Fig2_perf_per_rule}d). 

Interestingly, while arithmetic rules posed the greatest challenge for models to generalize, this difficulty did not arise from poor learning of the rules. Instead, it highlights the model's inability to generalize the arithmetic relation to new attributes. When evaluating test errors—the classification error on novel rows following the trained rules—arithmetic rules consistently had much lower errors than logical rules, which exhibited the highest errors across models (Supplementary Fig.1). 

Human performance on the original RPM task~\cite{raven1936mental} follows a hierarchy of difficulty, with progression rules the easiest and logical rules the hardest~\cite{forbes1964item,carpenter1990one,matzen1994error,matzen2010recreating}. Unlike models, humans do not receive explicit training on these rules before taking the task~\cite{te2001practice, flynn1987massive}. To evaluate whether models exhibit human-like reasoning, we compare their ability to generalize to novel rules against human performance. Among the models examined, only the SCL model replicates the human difficulty pattern across relation types (Fig.~\ref{fig:Fig2_perf_per_rule}d). To more rigorously test this alignment, we conducted additional human studies using the \textit{SimplifiedRPM} task, which allows a direct comparison of abstract relation generalization between humans and models.

To encompass a range of task difficulties, we sampled $130$ trials of the \textit{SimplifiedRPM} task, spanning $13$ relational rule pairs with $10$ questions per pair. We focused on rules defined by attributes that humans can readily discriminate, such as object numerosity, and avoided perceptually challenging attributes, such as subtle shade differences, to reduce perceptual difficulty for human participants since the focus of this study is reasoning and not visual perception. We recruited $n=25$ MTurk participants, each of whom answered $60$ randomly selected questions from this pool to minimize sequential learning effects. Participants had unlimited time to respond and were compensated based on accuracy. For each trial, we recorded both accuracy and response time (see Methods for details). Human participants exhibited a mean task error of $0.270\pm0.038$ (mean$\pm$s.e.m.), significantly lower than the chance level of $0.5$ (one-sample t-test, $t(12)=-5.70$, $p = 4.9 \times 10^{-5}$). The detailed $13$ rule pairs included in the human experiment and their corresponding human performance are in Supplementary Table 1.

We evaluated the generalization errors of the SCL model on the same \textit{SimplifiedRPM} trials used in human experiments. Specifically, we trained ten SCL models for each of the thirteen rule pairs, each using a different split of the forty rules in the dataset while ensuring that the target rule pair was always among the five held-out rules. We then measured the generalization error on the held-out pair using the same trials as those used in the human evaluation. The model achieved an average task error of $0.268\pm0.043$ (mean$\pm$s.e.m.), close to the human performance. Across the thirteen rule pairs, SCL models exhibited error patterns that correlated with those observed in human participants (Spearman’s rank correlation, $\rho=0.62$, $p=0.02$) (Fig.~\ref{fig:Fig2_perf_per_rule}e). 

Besides task errors, human response times—the total duration participants spent on each trial—were also strongly correlated with SCL generalization error across the thirteen rule pairs (Spearman’s rank correlation, $\rho=0.90$, $p=1.8\times10^{-5}$) (Fig.~\ref{fig:Fig2_perf_per_rule}f). This finding suggests that participants allocated more time to solving trials that the model found more challenging. In summary, although trained solely to classify relational rules without explicit human-like guidance, the SCL model successfully captures the human order of difficulty. In contrast, other models deviate from human performance patterns (Supplementary Fig.2). This alignment of SCL with human performance motivates further exploration of the mechanisms underlying how neural networks perform relational reasoning. 

\subsection*{Geometric Theory of Rule Manifolds Accurately Predicts Error in the \textit{SimplifiedRPM} Task}

To investigate how neural networks achieve relational reasoning and generalize their learned representations to unseen relational rules, we leveraged a geometrical theory of few-shot learning \cite{sorscher2022neural}. This theory provides a framework for linking errors in the Simplified RPM task to interpretable geometric properties of the relational representations.

\begin{figure}[ht]
\centering
\includegraphics[width=\linewidth]{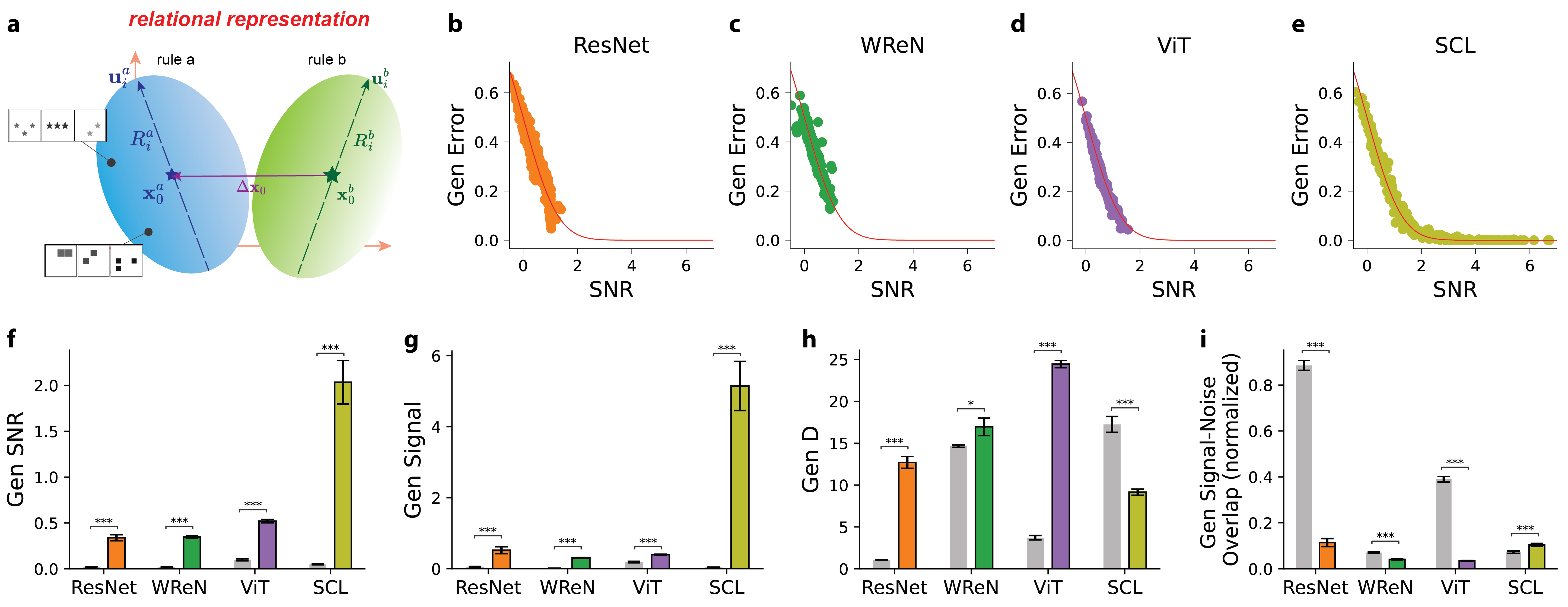}
\caption{Geometric Theory of Rule Manifolds Accurately Predicts Error in the \textit{SimplifiedRPM} Tasks.
    (a) Illustration of rule manifolds in the relational representation space. Orange axes denote representation dimensions. Each rule manifold consists of relational representations for rows that follow a specific rule. Two rule manifolds are shown: blue (rule a) and green (rule b) ellipsoids. The two black dots in rule a manifold represent example rows following rule A (e.g., constant shape). Each manifold is characterized by a centroid ($\mathbf{x}_{0}^{a}$ for rule a, blue star; $\mathbf{x}_{0}^{b}$ for rule b, green star), representing the mean position of all relational representations following that rule. Variation is characterized by principal axes ($\mathbf{u}_{i}^{a}$ for the blue rule manifold a and $\mathbf{u}_{i}^{b}$ for the green manifold b), shown as blue and green dashed arrows. Here, $i=1,.., N$, where $N$ is the relational representation dimension. The extent of variation along each axis is quantified by the corresponding radii $R_i^a$ and $R_i^b$. Purple line represents signal $\Delta\mathbf{x}_{0}$.  
    (b-e) Comparison of theoretical predictions (red lines) and empirical generalization errors (dots) across four models. Each dot represents a held-out rule pair, where the theoretical SNR predictions are computed using geometric measures (Equation~\ref{eq:SNR}), and the empirical generalization errors are averaged over 500 \textit{SimplifiedRPM} trials. Each plot includes rule pairs from 15 distinct held-out rule splits. Each split consists of five held-out rules, resulting in 20 rule pairs per split.
    (f-i) Generalization SNR, signal, D, and signal-noise overlap (normalized by signal magnitude) for model relational representations on the held-out rules. The gray bars represent the untrained models' representation at initialization, while the colored bars correspond to different models: orange for ResNet, green for WReN, purple for ViT, and yellow-green for SCL. Each bar shows the mean value averaged across 15 distinct held-out rule splits, with error bars indicating the s.e.m. Pairwise Wilcoxon tests assess whether the trained model terms differ from the untrained models. Asterisks indicate statistical significance: * ($p < 0.01$), *** ($p < 0.001$).}
\label{fig:Fig3_geometry_theory}
\end{figure}

We define \textit{rule manifolds} as the set of relational representations for rows that follow a specific rule (Fig.~\ref{fig:Fig3_geometry_theory}a). For a rule pair $(a,b)$, we consider the average generalization error $\varepsilon_{a}$ on the \textit{SimplifiedRPM} task across all trials where the sample and correct choice rows follow rule $a$, while the incorrect choice row follows rule $b$. Geometrically, $\varepsilon_{a}$ corresponds to the probability that a correct choice sampled from the rule a manifold is closer to an incorrect choice from the rule b manifold than to another correct choice from the rule a manifold.

Each rule manifold is defined by its centroid $\mathbf{x}_{0}$ (depicted as blue and green stars in Fig.~\ref{fig:Fig3_geometry_theory}a), which represents the mean position of all representations that follow a given rule. The spread of representations within a manifold is captured by a set of principal axes $\mathbf{u}_{i}$ (illustrated as blue and green dashed arrows in Fig.~\ref{fig:Fig3_geometry_theory}a), along which the variation is quantified by corresponding radii $R_i$. Here, $i=1,..., N$, where $N$ denotes the total dimensionality of the relational representation space. The overall magnitude of this variation is measured by the mean squared radius given by: $R^{2}\equiv\frac{1}{N}\sum_{i=1}^{N}R_{i}^{2}$. 

Although the rule manifolds may have complex shapes, the geometrical theory predicts that generalization error can be accurately estimated using only a few key geometric properties. Specifically, the generalization error, $\varepsilon_{a}$, for a given rule pair ($a$,$b$) is governed by the signal-to-noise ratio $\text{SNR}_{a}$, following: $\varepsilon_{a}=H\left(\text{SNR}_{a}\right)$. Here, $H\left(\cdot\right)$ is the Gaussian tail function defined as $H\left(x\right)=\intop_{x}^{\infty}dte^{-t^{2}/2}/\sqrt{2\pi}$. Higher $\text{SNR}_{a}$ values correspond to lower generalization errors. The dominant terms of $\text{SNR}_{a}$ are given by: 

\begin{equation}	                           
        \text{SNR}_{a}=\frac{\parallel\Delta\mathbf{x}_{0}\parallel^{2}+R_{b}^{2}R_{a}^{2}-1}{\sqrt{D_{a}^{-1}+\parallel\Delta\mathbf{x}_{0}\cdot\mathbf{U}_{a}\parallel^{2}+\parallel\Delta\mathbf{x}_{0}\cdot\mathbf{U}_{a}\parallel^{2}}}
	\label{eq:SNR} 
\end{equation}

The $\text{SNR}_{a}$ depends on four key interpretable geometric terms. The first is the \textit{signal}: ${\parallel\Delta\mathbf{x}_{0}\parallel^{2}=\parallel\mathbf{x}_{0}^{a}-\mathbf{x}_{0}^{b}\parallel^{2}/R_{a}^{2}}$, which measures the squared Euclidean distance between the centroids $(\mathbf{x}_{0}^{a}, \mathbf{x}_{0}^{b})$ of the two rule manifolds (depicted as purple line in Fig.~\ref{fig:Fig3_geometry_theory}a), normalized by the mean squared radius $R_{a}^{2}$ of rule $a$. A larger signal indicates greater separation between the rule $a$ and rule $b$ manifold centroids, leading to higher $\text{SNR}_{a}$ and lower generalization error. We denote $\Delta\mathbf{x}_{0}$ as the signal direction. The second term is the \textit{bias}, $R_{b}^{2}R_{a}^{-2}-1$, which captures the relative difference in the sizes of two rule manifolds. When rule manifold $a$ is larger than manifold $b$, the bias term is negative, predicting a lower $\text{SNR}_{a}$ for rule a. The third term, $D_{a}^{-1}$, represents the inverse of the dimensionality $D_a$ of the rule $a$ manifold. $D_a$, known as the ''Participation ratio''~\cite{gao2017theory}, is defined as $D_{a}\equiv\left(R_{a}^{2}\right)^{2}/\sum_{i=1}^{N}\left(R_{i}^{a}\right)^{4}$. $D_a$ quantifies the number of dimensions along which the manifold varies significantly. Higher-dimensional manifolds are preferred for minimizing generalization errors. The last term, \textit{signal-noise overlap}, is given by $\parallel\Delta\mathbf{x}_{0}\cdot\mathbf{U}_{a}\parallel^{2}$ and $\parallel\Delta\mathbf{x}_{0}\cdot\mathbf{U}_{b}\parallel^{2}$. Here, $\mathbf{U}_{a}\equiv\left[\mathbf{u}_{1}^{a}R_{1}^{a},...,\mathbf{u}_{N}^{a}R_{N}^{a}\right]$ and $\mathbf{U}_{b}\equiv\left[\mathbf{u}_{1}^{b}R_{1}^{b},...,\mathbf{u}_{N}^{b}R_{N}^{b}\right]$ are matrices of the manifold axes of variation. The signal-noise overlap quantifies the alignment between these axes of variation $(\mathbf{U}_{a}, \mathbf{U}_{b})$ and the signal direction $\Delta\mathbf{x}_{0}$. Smaller overlaps result in lower generalization errors. 

We estimate the SNR for each model by first extracting the relational representations of rows associated with each held-out rule, thereby forming the rule manifolds. We then computed the geometric properties of these manifolds and calculated the SNR for each held-out rule pair using Equation~\ref{eq:SNR}. Fig.~\ref{fig:Fig3_geometry_theory}b-e presents the computed SNR along with the empirically measured generalization error on the \textit{SimplifiedRPM} task, highlighting close alignments between theoretical predictions (red lines) and empirical results (individual points) across all four models. This alignment validates the geometrical theory as a principled framework for understanding abstract relation generalization through representation geometry.

The geometrical theory identifies three independent mechanisms that facilitate generalization: increasing signal, expanding dimensionality (D), and reducing signal-noise overlap. These mechanisms provide a structured framework for analyzing differences in generalization performance across models. Fig.~\ref{fig:Fig3_geometry_theory}f-i illustrates the geometry of held-out rule representations before (gray bars) and after (colored bars) training, highlighting key differences between models. For signal-noise overlap, we normalize the overlap by signal magnitude: $\frac{\parallel\Delta\mathbf{x}_{0}\cdot\mathbf{U}_{a}\parallel^{2}+\parallel\Delta\mathbf{x}_{0}\cdot\mathbf{U}_{b}\parallel^{2}}{\parallel\Delta\mathbf{x}_{0}\parallel^{2}}$. This normalization allows us to distinguish whether a high signal-noise overlap is due to a large signal magnitude or a genuine alignment between the manifold and the signal direction.

Across all models, training leads to increased SNR on held-out rules(Fig.~\ref{fig:Fig3_geometry_theory}f). Analyzing individual mechanisms reveals that SCL exhibits a pronounced increase in signal (Fig.~\ref{fig:Fig3_geometry_theory}g) ; however, this gain is accompanied by a decrease in dimensionality (D) and an increase in signal-noise overlap (Fig.\ref{fig:Fig3_geometry_theory}h,i). These opposing trends partially offset the benefits of higher signal, potentially limiting generalization. In contrast, ResNet, WReN, and ViT show more balanced geometric changes: although their signal gains are more modest, they are coupled with increased dimensionality and reduced overlap (Fig.~\ref{fig:Fig3_geometry_theory}g-i). These findings highlight that generalization generalization arises not from single geometric factor alone, but from the interplay of all geometric components identified by our theory.

\begin{figure}[ht]
\centering
\includegraphics[width=\linewidth]{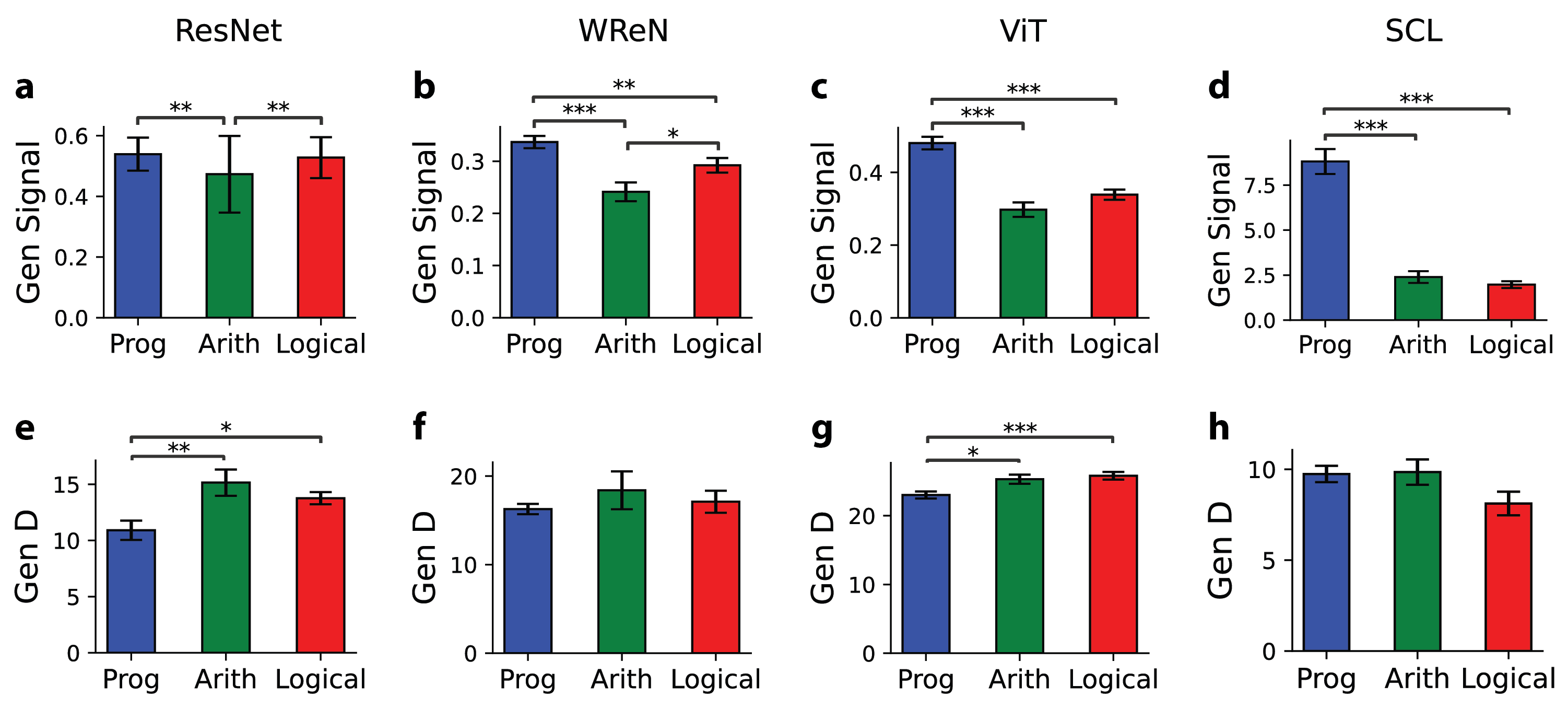}
\caption{Generalization error differences across abstract relation types were primarily influenced by signal rather than dimensionality.
    (a-d) Generalization signal and (e-h) dimensionality for different abstract relation types—progression, arithmetic, and logical—across four models. Results are averaged over 15 random splits of training and held-out rules. Each bar represents the average error for trials where the correct rule of the sample row belonged to each relation type, with error bars indicating the s.e.m. Statistical comparisons were conducted using the pairwise Wilcoxon test.}
\label{fig:Fig4_geometry_per_rule}
\end{figure}

We previously observed that generalization performance varied across abstract relation types (Fig.\ref{fig:Fig2_perf_per_rule}a–d). To investigate the geometric basis of these differences, we analyzed the signal, dimensionality, and signal-noise overlap of representations associated with each relation type. In the SCL model, which most closely mirrored human (Fig.\ref{fig:Fig2_perf_per_rule}d), differences in generalization were primarily governed by the signal: for example, the progression rule, which was easiest to generalize, exhibited the highest signal (Fig.\ref{fig:Fig4_geometry_per_rule}d). Similarly, in ResNet, WReN, and ViT, the arithmetic rules, with the highest generalization error, were associated with the lowest signal (Fig.\ref{fig:Fig4_geometry_per_rule}a-c). Dimensionality and overlap did not show a consistent relationship with relation type (Fig.\ref{fig:Fig4_geometry_per_rule}e–h; Supplementary Fig.3). Interestingly, arithmetic rules exhibited relatively higher dimensionality despite their poor generalization, suggesting that increased dimensionality, while potentially beneficial, is not by itself sufficient for abstraction. Together, these findings identify signal as the primary geometric correlate of generalization across abstract relations.

\subsection*{Layer-wise Geometric Analysis Reveals Model-Specific Generalization Strategies}

We observed that models exhibit distinct geometries in their relational representations. A key question is how these geometries emerge: Do they form abruptly at specific layers or gradually across the network? To address these questions, we examined the evolution of representational geometry across layers, aiming to pinpoint where abstract relations are extracted and how architecture shapes representation.

\begin{figure}[ht]
\centering
\includegraphics[width=\linewidth]{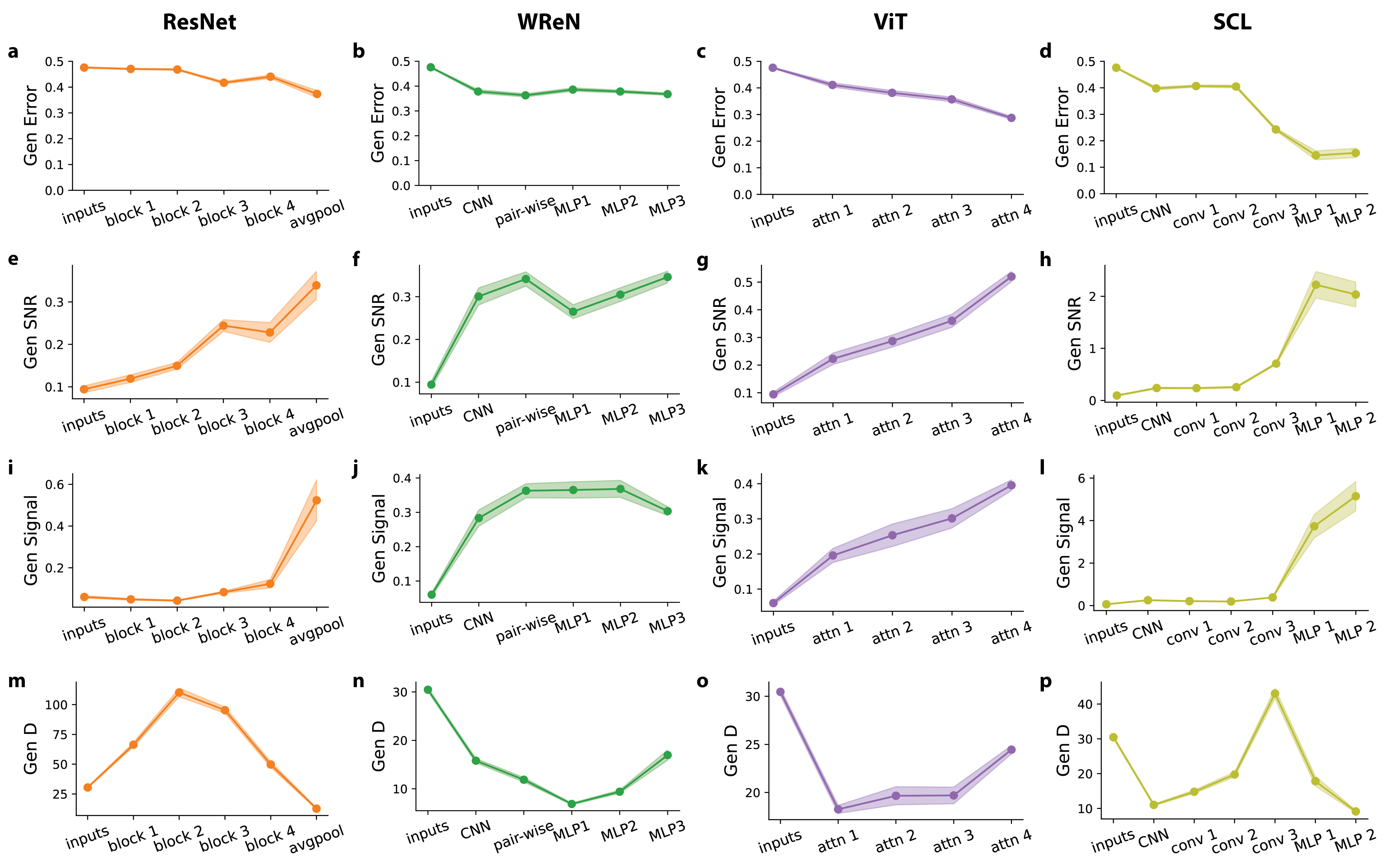}
\caption{Layer-wise Geometric Analysis Reveals Model-Specific Generalization Strategies.
    (a-d) Generalization error across layers for each model. The plotted values represent 15 distinct held-out rule splits, each containing five held-out rules, leading to 20 rule pairs per split. Error bars denote the s.e.m. across the 15 splits. Representations at each layer are randomly projected to a fixed dimensionality of $N=400$ before computing errors and performing geometric analysis. Generalization errors for each rule pair are evaluated using 500 \textit{SimplifiedRPM} trials.
    (e-h) Generalization signal-to-noise ratio (SNR) across layers.
    (i-l) Generalization signals across layers.
    (m-p) Generalization dimensionality (D) across layers.}
\label{fig:Fig5_layer_geometry}
\end{figure}

Fig.~\ref{fig:Fig5_layer_geometry}a-d shows each model's generalization error across layers. At the input stage, rows exhibit a generalization error of $0.475\pm0.002$, close to the $0.5$ chance level. This suggests the rules are not yet separated at the row pixel level. Across all four models, generalization errors decrease along the layers, demonstrating progressive refinement of relational representations. Fig.~\ref{fig:Fig5_layer_geometry}e-h shows generalization SNR across layers, highlighting distinct relational development patterns across architectures. ResNet-50 and ViT exhibit a gradual increase in SNR, reflecting a steady accumulation of relational information (Fig.~\ref{fig:Fig5_layer_geometry}e,g). In contrast, WReN plateaus early at the pairwise aggregation layer, with later MLP layers failing to improve further generalization SNR (Fig.~\ref{fig:Fig5_layer_geometry}f). In the SCL model—which achieved the best generalization performance—we observed a sharp increase in generalization SNR at the first MLP layer, but decreased slightly at the second (final) layer  (Fig.~\ref{fig:Fig5_layer_geometry}h). 

To gain deeper insights into how relational rule separability evolves across network layers, we decompose SNR into the dynamics of its geometric terms at each layer. While signal generally increases across layers, its layerwise trend does not always follow that of SNR. While SNR steadily rises in ResNet, the signal remains relatively stable until a sharp increase in the final average pooling layer (Fig.~\ref{fig:Fig5_layer_geometry}i). In WReN, the signal plateaus at the pairwise aggregation layer and even decreases in the final MLP layer (Fig.~\ref{fig:Fig5_layer_geometry}j). In ViT, the signal progressively increases across layers (Fig.~\ref{fig:Fig5_layer_geometry}k). In SCL, the signal only begins to rise sharply in the first MLP layer, whereas SNR increases earlier in the last convolutional layer. Furthermore, while the signal rises in the second MLP layer, SNR unexpectedly decreases (Fig.~\ref{fig:Fig5_layer_geometry}l).  

The observed misalignment suggests that signal amplification alone is not sufficient to explain rule separation across layers. To further investigate, we examine the evolution of dimensionality (D) (Fig.~\ref{fig:Fig5_layer_geometry}m-p). Interestingly, unlike the signal, D exhibits a non-monotonic trajectory across layers. ResNet-50 initially expands D in early layers before compressing in later layers (Fig.~\ref{fig:Fig5_layer_geometry}m). WReN follows a different pattern, initially compressing D through pairwise aggregation and early MLP layers before expanding it in later MLP layers (Fig.~\ref{fig:Fig5_layer_geometry}n). ViT maintains a consistently low D throughout, with only a slight increase in the final attention block (Fig.~\ref{fig:Fig5_layer_geometry}o). SCL increases D through convolutional layers but then sharply compresses it in MLP layers (Fig.~\ref{fig:Fig5_layer_geometry}p). These variations in D help explain the gaps observed between signal and SNR evolution through layers. In ResNet-50,  the initial expansion of D accounts for the early rise in SNR, which occurs before the signal increases. In SCL, the sharp compression of D in the last MLP layer explains why SNR does not increase alongside the signal. 

Across different models, our findings on layerwise geometry suggest that relational abstraction is not a uniform process; each model employs a distinct strategy. The SCL model achieves the best generalization performance by maximizing signal amplification, especially in MLP1, outperforming all other models. In contrast, ResNet-50 and ViT progressively refine relational representations, whereas WReN reaches an early plateau, indicating a bottleneck in relational generalization. While the previous section showed weaker correlations between D and rule separation across abstract relation types, the layerwise analysis reveals a more dynamic interaction. Specifically, D plays a crucial role in shaping relational representations across model layers by refining the effectiveness of signal amplification in driving their emergence.

\subsection*{Structured Compression Drives the Trade-off Between Generalization Signal and Dimensionality}

\begin{figure}[ht]
\centering
\includegraphics[width=\linewidth]{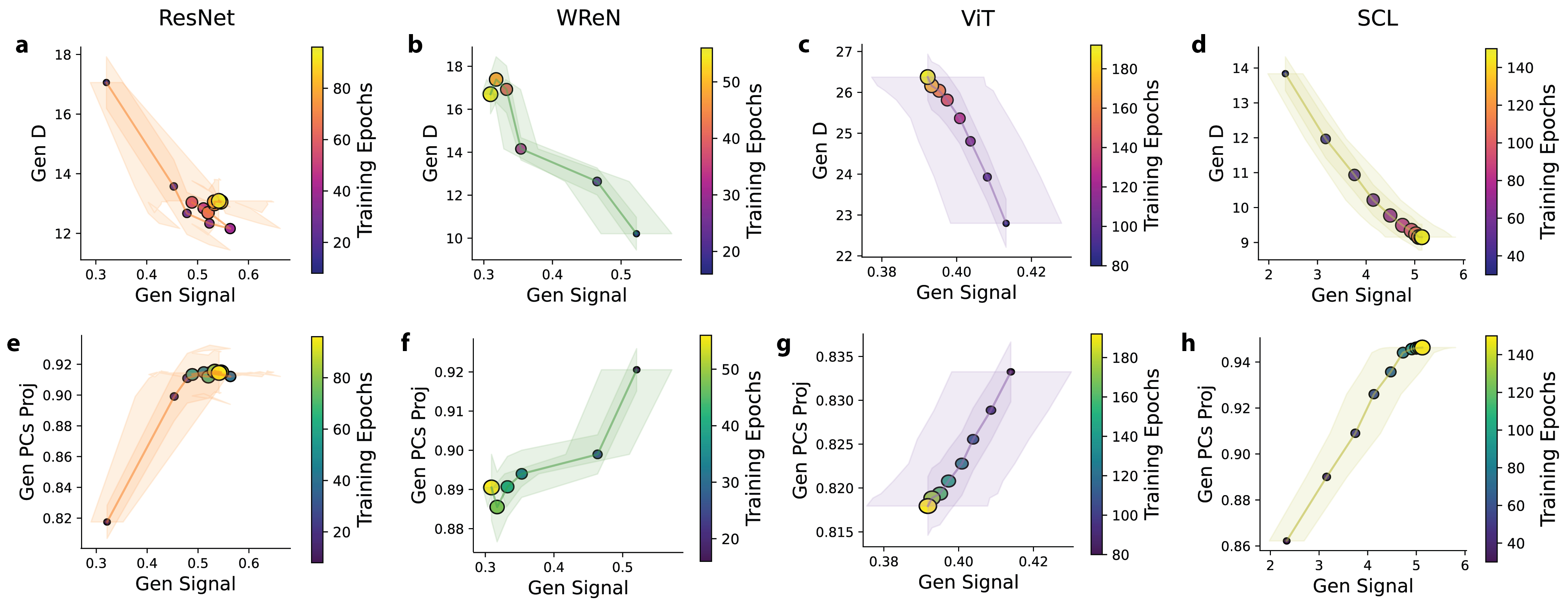}
\caption{Structured Compression Drives the Trade-off Between Generalization Signal and Dimensionality.
    (a-d) Generalization signal vs. dimensionality (D) across training epochs for all models. Each point shows average signal and D over 15 held-out splits (five rules per split). Dot size and color indicate training epoch; later epochs are larger, brighter. Spearman’s rank correlation between the generalization signal and D reveals a strong negative correlation across the 15 rule splits for each model (ResNet: \( \rho = -0.12\pm0.12 \); WReN: \( \rho = -0.41 \pm 0.11 \); ViT: \( \rho = -0.37 \pm 0.19 \); SCL: \( \rho = -0.77 \pm 0.05 \); mean \( \pm \) s.e.m.).  This effect was statistically significant, as confirmed by Stouffer’s test (ResNet: $Z = 3.41$, $p = 3.2 \times 10^{-4}$; WReN: $Z = 2.38$, $p = 8.5 \times 10^{-3}$; ViT: $Z = 14.76$, $p = 1.1 \times 10^{-49}$; SCL: $Z = 13.04$, $p = 3.5 \times 10^{-39}$). 
    (e-h) Alignment of held-out rule manifolds with the low-dimensional subspace of training rules across training epochs for all four models. Each point shows the variance explained by the top 35 principal components of training rule representations for a held-out rule manifold, averaged over 15 splits. The size and color of the dots indicate the training epoch, with later epochs depicted by larger, brighter dots. Spearman’s rank correlation between the generalization signal and subspace alignment shows a strong positive correlation across the 15 rule splits for each model (ResNet: \( \rho = 0.01 \pm 0.13 \); WReN: \( \rho = 0.29 \pm 0.11 \); ViT: \( \rho = 0.38 \pm 0.19 \); SCL: \( \rho = 0.90 \pm 0.04 \); mean \( \pm \) s.e.m.). This effect was statistically significant in all four models except WReN, as confirmed by Stouffer’s test (ResNet: $Z = 3.35$, $p = 3.9 \times 10^{-4}$; WReN: $Z = 0.37$, $p = 0.35$; ViT: $Z = 12.09$, $p = 6.0 \times 10^{-34}$; SCL: $Z = 19.54$, $p = 2.5 \times 10^{-85}$).We begin plotting from the training epoch, where the structured low-dimensional representation of the training rule first emerges, indicated by the stabilization of its relational representation’s dimensionality. The full trajectory of the generalization signal, D, and alignment is presented in Supplementary Fig.8.}
\label{fig:Fig6_signalD_tradeoff}
\end{figure}

The geometrical theory suggests that high signal and dimensionality (D) benefit generalization. However, we observed that in model layers where significant signal increases occur, D tends to decrease (e.g., Fig.~\ref{fig:Fig5_layer_geometry}i, m for ResNet-50; Fig.~\ref{fig:Fig5_layer_geometry}l, p for SCL). This relationship between the generalization signal and D occurs not only across model layers but also throughout training. In all four models, signal and D exhibit a consistent anti-correlation, with signal increases accompanied by decreases in D (Fig.~\ref{fig:Fig6_signalD_tradeoff}a-d). Intuitively, signal and D are two independent geometric properties. For example, the centroids of two rule manifolds can remain fixed, preserving the signal, while their shapes evolve, changing D. This raises a key question: Why does this trade-off between the generalization signal and D consistently emerge?

We propose that the trade-off between the generalization signal and D arises from a structured compression effect. All class manifolds—including those for held-out rules—gradually align with a low-dimensional subspace defined by the principal components of the training rules. Across models, the representations of the 35 training rules consistently occupy a low-dimensional space. Notably, we observe a drop in variance explained by the top principal components (PCs) at exactly $35$, matching the number of training rules (Supplementary Fig.4). We hypothesize that the network amplifies the generalization signal by aligning all representations, including those of held-out classes, within this low-dimensional space. 

To test this hypothesis, we measured the fraction of variance explained by the top $35$ principal components (PCs) of the training rules for each held-out rule manifold. Higher values indicate more substantial alignment with the low-dimensional subspace of training rules. We observed that as the network signal increases, this alignment becomes more pronounced across models (Fig.~\ref{fig:Fig6_signalD_tradeoff}e-h).

Our findings suggest that the observed signal-D trade-off is not incidental but a consequence of structured compression. As networks enhance the generalization signal, they constrain representations into a low-dimensional subspace, prioritizing inter-class signal at the expense of dimensionality. This effect is a general mechanism across different architectures.

\subsection*{Theory-Grounded SNR Loss Yields Geometry-Balanced Representations}

Can the geometry predicted by the theory directly guide neural network training? We hypothesize that enforcing this geometry provides a principled approach to learning relational representations. To test this hypothesis, we introduce the SNR loss—an SNR-based objective function that explicitly enforces the predicted geometry.  

Specifically, the SNR loss maximizes the SNR for the training rule pairs, reinforcing the optimal representation geometry. We sample $P$ rows from $m$ relational rules for each input batch and process them through the model to obtain high-dimensional relational representations. The SNR loss is then formulated as: 
\begin{equation}
	l_{SNR}=\sum_{(a,b)}\exp\left(-SNR_{a}\right)
	\label{eq:SNR_loss} 
\end{equation}
Here, $SNR_{a}$ represents the estimated SNR (~Equation~\ref{eq:SNR}) for the ordered rule pair $(a,b)$, averaged across all possible rule pairs in the batch. 

\begin{figure}[ht]
\centering
\includegraphics[width=\linewidth]{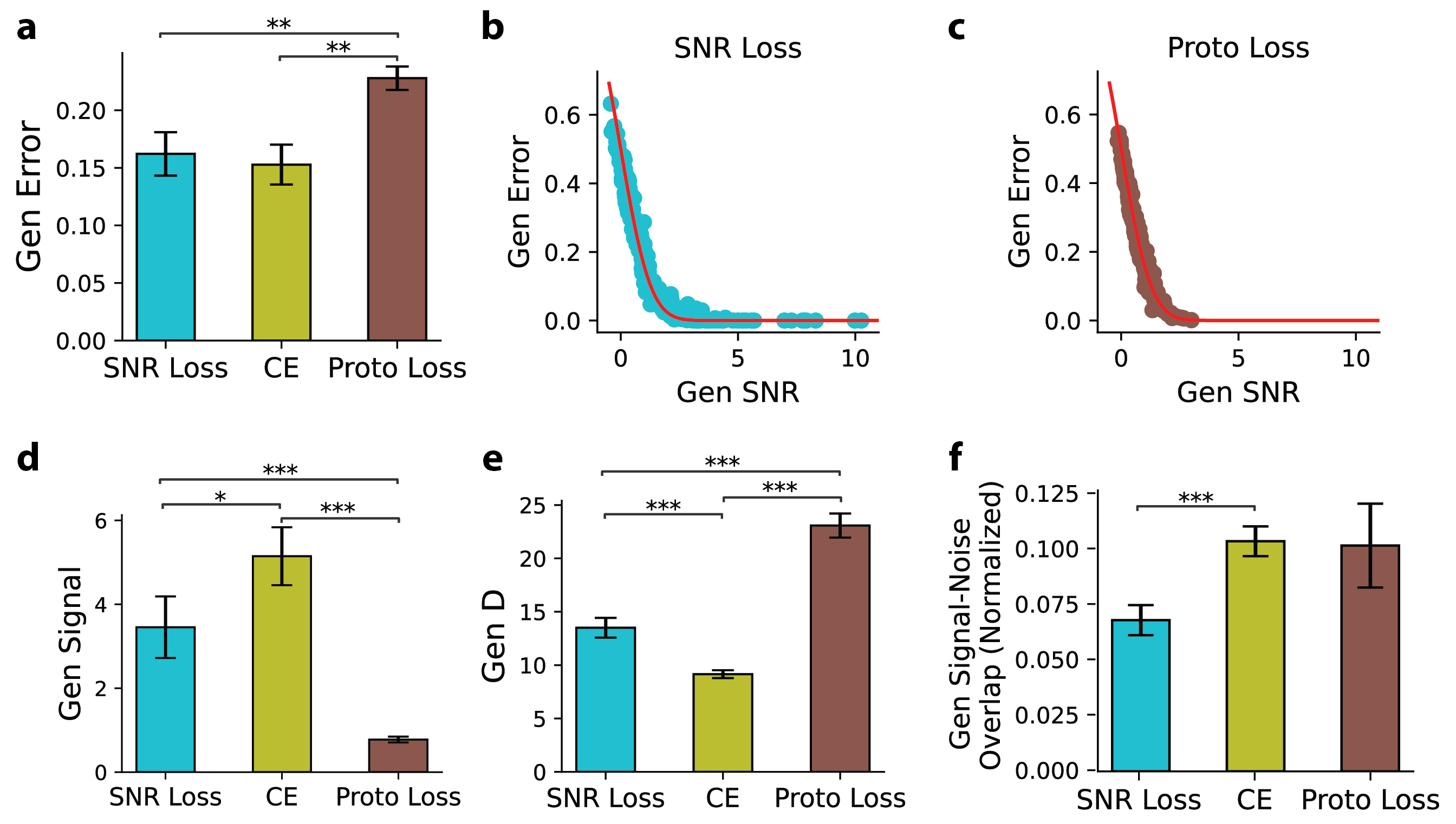}
\caption{Theory-Grounded SNR Loss Yields Geometry-Balanced Representations.
    (a) Generalization error for models trained with SNR loss, cross-entropy (CE) loss, and Prototypical (Proto) loss, averaged over 15 rule splits. Error bars denote s.e.m. Pairwise Wilcoxon tests confirm Prototypical achieves the highest error ($p$-values: SNR Loss vs. Proto Loss, $5.1 \times 10^{-3}$; CE vs. Proto Loss, $1.0 \times 10^{-3}$). 
    (b,c) Comparison of theoretical predictions (red lines) and empirical generalization errors (dots) for the SNR and Prototypical Loss. Each dot represents a held-out rule pair.
    (d-f) Generalization signal, dimensionality (D), and signal-noise overlap (normalized by signal magnitude) for models trained with SNR, CE, and Proto Loss. Each bar represents the mean value across 15 rule splits, with error bars indicating the s.e.m. Pairwise Wilcoxon tests assess the statistical significance of differences between loss functions. Asterisks indicate statistical significance: * ($p < 0.05$), ** ($p < 0.01$), *** ($p < 0.001$).}
\label{fig:Fig7_SNRloss}
\end{figure}

To assess the effectiveness of the proposed SNR loss, we trained the top-performing SCL model using this objective and evaluated its generalization across the same $15$ rule splits previously used for training with cross-entropy loss. The model effectively optimized the training rule representations, achieving an average SNR of $9.66\pm0.22$ (mean$\pm$s.e.m.) across training rule pairs, as evaluated on novel test rows of the training rules. More importantly, the model effectively separated held-out rules, yielding an averaged generalization error of $0.162\pm0.018$ (mean$\pm$s.e.m.) (Fig.~\ref{fig:Fig7_SNRloss}a). This performance was statistically indistinguishable from models trained with cross-entropy (CE) loss using one-hot rule labels (Wilcoxon signed-rank test: statistic = $48.0$, $p=0.26$, n.s.), confirming that SNR loss achieves competitive generalization while structuring learned representations according to theoretical predictions.

To further evaluate the efficacy of SNR loss, we compared its performance against Prototypical loss, a widely used loss function designed to learn structured representations that generalize to unseen classes \cite{snell2017prototypical}. Prototypical loss operates by embedding inputs into a high-dimensional representation space and forming class prototypes, defined as the centroids of the representations corresponding to each class. The loss function then encourages representations of samples from the same class to cluster around their respective prototypes. Additional details on Prototypical loss are provided in the Supplementary Text. We trained the SCL model using Prototypical loss on the same $15$ rule splits. The resulting model exhibited a generalization error of $0.227\pm0.010$, significantly higher than that obtained with SNR loss (Fig.~\ref{fig:Fig7_SNRloss}a). These findings suggest that while both loss functions promote structured representations, SNR loss is more effective in shaping a geometric representation that enhances generalization.

Notably, the generalization errors of models trained with both SNR loss and Proto loss, similar to our previous observations with cross-entropy loss (Fig.~\ref{fig:Fig3_geometry_theory}e), are also accurately predicted by the geometrical theory (Fig.~\ref{fig:Fig7_SNRloss}b,c). Compared to cross-entropy (CE) and Prototypical loss, SNR loss is uniquely grounded in geometric theory, explicitly accounting for key representation geometries, including signal strength, dimensionality, and signal-noise overlap. This theoretical foundation raises the question of whether SNR loss yields a more advantageous representational geometry relative to other loss functions. To investigate this, we analyzed the representation geometries induced by different loss functions. Our findings show that representations learned with SNR loss have higher dimensionality than those trained with CE loss (Fig.~\ref{fig:Fig7_SNRloss}e), addressing the low-dimensionality issue in CE training while maintaining relatively high signal strength (Fig.~\ref{fig:Fig7_SNRloss}d). Furthermore, SNR loss reduces signal-noise overlap, which is beneficial for generalization (Fig.~\ref{fig:Fig7_SNRloss}f). Although Prototypical loss also encourages high-dimensional representations, it results in the lowest signal, contributing to its worst generalization performance (Fig.~\ref{fig:Fig7_SNRloss}d, e). These findings highlight that SNR loss shapes the representations of held-out rules in a way that explicitly adheres to geometric principles.

\section*{Discussion}

A longstanding challenge in both cognitive science and artificial intelligence is to understand how neural systems generalize abstract relations beyond the rules they were trained on. Here, we take a first step toward addressing this question by introducing a principled geometric framework that links generalization performance to the structure of rule representation. We hypothesize that networks trained to perform reasoning tasks develop structured internal representations even for unseen rules, and these representations form distinct rule manifolds. Building on geometric theory originally developed for object recognition~\cite{sorscher2022neural}, we propose that generalization performance in visual reasoning can be explained by key geometric properties: manifold radii, dimensionality, the distance between manifold centroids, and the overlap between these distances and their subspaces of variation.

To systematically investigate this hypothesis, we introduced \textit{SimplifiedRPM},  a curated dataset of Raven’s Progressive Matrices tasks that enables systematic evaluation of generalization in both humans and neural networks. We tested a range of models (ResNet~\cite{he2016deep}, WReN~\cite{barrett2018measuring}, ViT~\cite{dosovitskiy2020image}, and SCL~\cite{wu2020scattering}), as well as human participants on the same dataset. Among these, SCL achieved the strongest generalization and most closely mirrored human difficulty rankings across abstract relations. The current work expands on the initial probing of the SCL architecture by providing a quantitative account of compositional generalization and 
by identifying specific geometric properties that predict generalization performance.

Our analysis revealed that representation geometry accurately predicts generalization ability, and that different models strike different balances between signal and dimensionality. Critically, we observed that not all relational rules are equally generalizable. The best-performing model, SCL, struggled with logic-based rules, which humans also found challenging. Other models, despite high test accuracy on arithmetic rules, failed to generalize them, highlighting a disconnect between testing evaluation and the deeper relational understanding necessary for generalization. Our representation analysis revealed that signal, more than dimensionality or signal-noise overlap, governs these differences across rules. Building on these insights, we propose SNRloss, a theoretically grounded objective function that explicitly optimizes representation geometry. This novel loss function provides adequate training and generalization capabilities. 

Layer-wise analyses revealed distinct trajectories in the emergence of relational representations across architectures. In SCL, relational structure emerges abruptly at the first multilayer perceptron (MLP) layer following the convolutional blocks, where global relational information likely becomes accessible through integration across multiple panels. However, this increase does not persist, and the SNR decreased slightly in the subsequent MLP layer. In contrast, Vision Transformers (ViTs) exhibit a more gradual and continuous refinement of relational structure across layers. An intriguing question is whether high-level reasoning operates through discrete, stage-like computations as in SCL, or emerges gradually as in ViTs.

Our findings also refine theoretical predictions about dimensionality. While higher dimensionality can support richer representations and better generalization~\cite{sorscher2022neural}, prior works on object classification also suggest advantages for low-dimensional manifolds in enabling the classification of a greater number of objects~\cite{cohen2020separability, chung2018classification}. This contrast implies that dimensionality should not be universally maximized or minimized but instead tuned to task demands. Our findings extend this idea by showing that generalization depends not only on dimensionality but mostly on its interaction with pairwise manifold signals. Specifically, we demonstrate that deep networks systematically compress representations into structured, low-dimensional spaces—even for unseen rules. Moreover, this phenomenon is consistently observed across diverse architectures (e.g., ResNet, SCL), indicating that it is a fundamental learning property rather than an artifact of any specific architecture. These findings open new avenues to explicitly leverage this signal versus dimensionality trade-off to improve generalization.

While prior work with Bayesian models accounts for human difficulty patterns across abstract relations by explicitly incorporating prior probabilities for each rule type \cite{little2012bayesian}, our findings demonstrate that the SCL model—trained end-to-end without any rule-type-specific assumptions—nonetheless captures human-like patterns of relational difficulty. Furthermore, SCL's generalization errors closely correlate with human response times. While these correlations certainly do not imply that SCL reasons like humans, our geometric analysis provides a principled foundation for future investigations into the cognitive factors underlying this correlation. The SCL network parallels the brain's hierarchy: its CNN-based feature extractor mimics the ventral visual cortex, capturing basic features like shape \cite{freedman2003comparison, yamane2006representation,kreiman2021, nayebi2024neural}, while its later reasoning module reflects higher cortical areas supporting relational reasoning \cite{davis2017concrete, chafee2007representing, summerfield2020structure, nelli2023neural}. We propose that the brain encodes rules of input rows through distinct, rule-specific neural manifolds. Our observation of layer-wise changes in the geometry of these rule manifolds offers hypotheses into how neural representations evolve across brain regions~\cite{mcintosh2016deep}. According to the SCL model, early visual areas exhibit high representational dimensionality (D) but low signal, whereas later regions show larger signals with reduced dimensionality. These predictions can be empirically tested by measuring neural activity during abstract reasoning tasks~\cite{courellis2024abstract, krawczyk2012cognition, langdon2023unifying}. 

This study aimed to study the generalization of abstract relations. While the humans in our experiment were not explicitly trained on the rules, their reasoning abilities could be considered "pre-trained" through life experiences, as they may have previously encountered similar rules. In contrast, neural networks were trained under strictly controlled rule splits, ensuring they had no prior exposure to the rules evaluated for generalization. To address this discrepancy, a crucial next step is to develop a new benchmark for human reasoning that incorporates abstract relations unlikely to have been encountered. Such a benchmark would disentangle relational generalization from prior experience, providing a clearer view of their distinct roles in human abstract reasoning. This approach enables a direct comparison between humans and neural networks, offering deeper insights into their generalization mechanisms.

In summary, our findings not only advance our understanding of relational reasoning in neural networks but also provide testable hypotheses on how geometric principles may underlie abstract reasoning in the brain. Looking forward, our approach opens exciting opportunities for extending geometric analysis to a broader range of cognitive tasks, including real-world visual reasoning benchmarks and cross-modal linguistic challenges.

\section*{Methods}

\subsection*{\textit{SimplifiedRPM} Dataset}
The Raven’s Progressive Matrices (RPM) task~\cite{raven1936mental} is a widely used assessment of abstract relational reasoning, commonly applied in fields such as cognitive psychology and AI evaluation. Each RPM trial presents a 3×3 grid with the bottom-right panel missing. The subject must select the correct answer from eight options to complete the missing panel, ensuring that all three rows or columns in the matrix follow a consistent underlying relational rule.

Inspired by the RPM task, the \textit{SimplifiedRPM} dataset was specifically designed to assess a model’s ability to generalize abstract relations across different object attributes (Fig.~\ref{fig:Fig1_RPMtask_and_perf}). Each row in the dataset follows a unique relational rule that determines how an object’s attribute values change across three panels (Supplementary Fig.5a). The dataset consists of $40$ relational rules, each derived by applying an abstract relation to a specific object attribute. These abstract relations fall into three categories: progression (constant, progression $\pm2$, progression $\pm1$), arithmetic (addition, subtraction), and logical (AND, OR, XOR). A detailed description of the abstract relations is provided in Supplementary Table 2.

Each rule was applied to one of four object attributes: shape, size, shade, or number/position (Supplementary Fig.5). Number and position were grouped as a single attribute to avoid ambiguity. For example, a "constant position" rule inherently fixes the object count, enforcing the constant number rule and conflicting with our requirement that each row in the task follows a unique relational rule. To ensure consistency with prior benchmarks, the dataset adopts the same object definitions and attribute specifications as the PGM dataset \cite{barrett2018measuring}. Shape is selected from seven geometric forms: circle, triangle, square, pentagon, hexagon, octagon, and star. The objects' sizes are represented by their bounding boxes in the panel, with heights ranging from $20$ to $40$ pixels in ten evenly spaced increments. Shade is represented as a grayscale intensity, ranging from $[230, 230, 230]$ (close to white) to $[0, 0, 0]$ (black) in RGB values, with $10$ evenly spaced increments. The number attribute specifies the number of objects per panel, ranging from $1$ to $9$. Position refers to the spatial arrangement of objects. Each panel has a total size of $[160, 160]$, and each object occupies a maximum bounding box of $[40, 40]$. There are nine possible object locations, each defined by its centroid coordinates: $\{(20,20),(20,60),(20,100),(60,20),(60,60),(60,100),(100,20),$
$(100,60),(100,100)\}$. Consequently, each row has dimensions $3\times160\times160$. When presented, the three panels are arranged horizontally. 

Dataset generation followed a structured pipeline to ensure consistency and reproducibility. Each row was constructed to enforce a unique rule from our predefined set, eliminating ambiguity. To achieve this, we first sampled attribute values from valid options that conformed to the designated rule. Next, we randomly assigned values to the remaining attributes and resampled if a sampled value inadvertently conformed to an unintended rule. Once attribute values were finalized, they uniquely defined an input row. The corresponding row input was then generated by placing objects with the specified attributes in their designated positions. To maintain dataset balance, we generated 10,000 training rows, 1,000 validation rows, and 1,000 test rows per rule, yielding a total of 48,000 unique rows across all rules (40 rules $\times$ 1,200 samples). Example rows in the dataset are provided in Supplementary Fig.5b-d.

\subsection*{Neural Network Training}
To evaluate a model’s ability to generalize abstract relations to unseen object attributes, we implement a rule-splitting strategy in which a subset of rules is withheld during training. Specifically, we randomly select five rules to be held out, ensuring that each abstract relation appears at most once in this set. The remaining $35$ rules are used for model training. After training, the model is evaluated based on its ability to infer the held-out rules. Supplementary Fig.5a provides an example of this rule split, with held-out rules highlighted in yellow and training rules in blue.

During training, models are optimized using supervised learning with a cross-entropy loss function. The models are trained on 10,000 examples per rule across $35$ training rules. Test error (Fig.~\ref{fig:Fig1_RPMtask_and_perf}c) is assessed based on classification accuracy using $1,000$ test examples per rule. These test examples differ from the $10,000$ training examples used during model training.

After training, we evaluate the model’s ability to infer previously unseen, held-out rules. For each pair of the five held-out rules (20 possible rule pairs), we generate 500 \textit{SimplifiedRPM} trials. Each trial is constructed by randomly sampling two rows from the first rule—one as the sample row and the other as the correct choice. Then, we sample a row from the second rule as the incorrect choice. Finally, we ensure all 500 trials per rule pair are unique. 

To evaluate model performance, we compute the Euclidean distance between the sample row and each of the two choice rows. A trial is considered correct if the correct choice row is closer to the sample row than the incorrect one. The model’s generalization error is then quantified as the proportion of incorrect trials averaged across all $500$ trials for each of the $20$ rule pairs (Fig.~\ref{fig:Fig1_RPMtask_and_perf}e). 

We trained four architectures: ResNet-50, WReN, ViT, and SCL. ResNet-50~\cite{he2016deep} is a convolutional neural network with residual connections. We employed the standard ResNet-50 architecture without modification. WReN~\cite{barrett2018measuring} explicitly models pairwise relations between objects and has been previously used for RPM tasks. We optimized the hyperparameters of its relation module and fully connected layers. ViT~\cite{dosovitskiy2020image} is a transformer-based model that operates on image patches as input tokens and captures global dependencies via self-attention. We split each of the three panels in a row into a $3\times3$ grid, resulting in $27$ patches per row. We optimized key hyperparameters, including the number of layers, attention heads, and hidden dimensions. Finally, SCL~\cite{wu2020scattering} was specifically designed for the RPM task, and we used the original architecture without modifications (Supplementary Fig.6). Detailed model configurations, including layer-wise specifications and hyperparameter tuning, are provided in Supplementary Text.

Training was conducted using stochastic gradient descent with a learning rate of $0.1$, momentum of $0.9$, and weight decay of $5 \times 10^{-4}$. To facilitate stable convergence, we applied a learning rate warmup for the first two epochs, followed by a plateau-based scheduler that reduced the learning rate by $0.2$ whenever the validation loss plateaued. Training continued until the classification errors across $1,000$ validation rows for each of the $35$ training rules stabilized. The final training durations for each model were: ResNet-50 ($100$ epochs), WReN ($60$ epochs), ViT ($200$ epochs), and SCL ($150$ epochs). All models were trained using a batch size of $512$. 

\subsection*{Human Experiment}
Twenty-five participants were recruited via Amazon Mechanical Turk (MTurk) to evaluate performance on the \textit{SimplifiedRPM} dataset. To ensure data reliability, we set strict eligibility criteria, requiring an approval rating above $95\%$ and restricting participation to Masters-qualified workers. All participants provided informed consent, and the study protocol was approved by the Institutional Review Board at Children’s Hospital, Harvard Medical School. Compensation was performance-based, with participants earning more for higher accuracy. All methods were performed in accordance with the relevant guidelines and regulations.

Before the main experiment, participants read written instructions describing the task and completed three practice trials that introduced relational rules different from the $13$ rule pairs used in the main experiment. Participants received feedback and explanations of the underlying rules during these practice trials. In the main experiment, each participant completed 60 trials randomly selected from a larger pool of $130$ trials ($10$ trials per rule pair), encompassing $13$ distinct rule pairs of varying relational complexity (Supplementary Table 1). Participants were shown images like those in Fig.~\ref{fig:Fig1_RPMtask_and_perf}a. The images remained on the screen until the participants made a choice. Participants responded by clicking on one of two choice rows and were allowed unlimited time to provide their answers. No performance feedback was given during the main experiment. Additional information, including illustrative example trials, can be found in Supplementary Fig.7.

\bibliography{sample}

\section*{Funding} 
This work was supported by the National Institutes of Health grant R01EY026025, the Swartz Foundation, the Office of Naval Research grant No. N0014-23-1-2051, and the Kempner Institute for the Study of Natural and Artificial Intelligence at Harvard University.

\section*{Author contributions}
J.S., G.K., and H.S. designed research; J.S. performed research; J.S. analyzed data; J.S., G.K., and H.S. wrote the paper.

\section*{Data availability}
The code and dataset used in the study are publically available on Github: https://github.com/ShaneShang/SimplifiedRPM-Visual-Reasoning. 

\section*{Competing interests} 
The authors declare no competing interests.

\clearpage
\appendix
\clearpage
\begin{center}
    \LARGE\textbf{Supplementary Information}\\[1.5ex]
    \Large Unraveling the Geometry of Visual Relational Reasoning\\[1.5ex]
    \large Jiaqi Shang, Gabriel Kreiman, and Haim Sompolinsky\\[1ex]
    \normalsize\textit{*Correspondence: hsompolinsky@mcb.harvard.edu}
\end{center}
\vspace{2em}
\date{}

\renewcommand{\thefigure}{S\arabic{figure}}
\setcounter{figure}{0}
\renewcommand{\thetable}{S\arabic{table}}
\setcounter{table}{0}

\section{Supplementary Methods}
\subsection{Model Architecture and Hyperparameter Tuning}
We used the standard ResNet-50 architecture described by \cite{he2016deep} without any modifications. The architecture comprises residual blocks containing convolutional layers followed by batch normalization and ReLU activation. Specifically, the model begins with an initial convolutional layer, followed by four residual blocks, and concludes with a global average pooling layer and a fully connected layer for classification. To analyze the evolution of layer-wise geometry, we extracted features from the outputs of each residual block (Block 1–4) and the final output of the average pooling layer (avgpool).

The Wild Relation Network (WReN) model captures and integrates pairwise relational information from the three panels in an input row. Each panel is first processed through a shared four-layer convolutional block to generate an individual panel embedding. Then, embeddings from each pair of panels are concatenated and passed through a multi-layer perceptron (MLP). The resulting pairwise features are summed to form an aggregated relational representation, which is further refined through three additional MLP layers. For layer-wise geometric analysis, we extract representations from the convolutional block output (CNN), the aggregated pairwise features (Pairwise), and the three successive MLP layers (MLP1–3).

For hyperparameter tuning of WReN, we performed a grid search over the following parameters: (1) hidden dimension of the pairwise relational module $\{128, 256, 512\}$; (2) number of layers in the pairwise relational module $\{1, 2, 3, 4\}$; (3) hidden dimension of the classification MLP $\{64, 128, 256\}$; and (4) number of layers in the classification MLP $\{1, 2, 3, 4\}$. The optimal configuration, minimizing test error on three held-out rule splits, was found to be a pairwise relational module with $2$ layers of $512$ dimensions and a classification MLP with $3$ layers of $256$ dimensions.

The Vision Transformer (ViT) architecture comprises multiple layers of multi-headed self-attention (MSA). In our implementation, we utilize a four-layer Transformer model, each containing $12$ attention heads and a hidden dimension of $96$. The model processes input row patches, embedding them into token representations with a learnable class token prepended to the sequence. These representations are then sequentially processed through self-attention layers to capture global contextual dependencies. For layer-wise geometric analysis, we extract the outputs from each attention layer (Attn1-4).

For hyperparameter tuning of ViT, we performed a grid search over the following parameters: (1) number of layers $\{4,8,12\}$; (2) number of attention heads $\{4,8,12,16\}$; (3) hidden dimension of the self-attention layers $\{48,96,192,384,768\}$. The optimal configuration, determined by minimizing test error, was found to be $4$ layers, each with $12$ heads and a hidden dimension of $96$.

The Scattering Compositional Learner (SCL) model~\cite{wu2020scattering} is a hierarchical architecture designed explicitly for RPM tasks (figure~\ref{fig:SuppFig6_SCL_architecture}). Each input panel is first processed through a shared convolutional block, producing an individual panel embedding. These embeddings are then processed through three 1-D convolutional layers that operate across the three panels. Crucially, the same filter is applied consistently across the panel-wise feature dimensions, ensuring that abstract relations are detected in an attribute-invariant manner. Finally, the output is further refined using two MLP layers. For layer-wise geometric analysis, we extract representations from the convolutional block (CNN), the three 1-D convolutional layers (conv1-3), and the two MLP layers (MLP1–2). 

For hyperparameter tuning of SCL, we performed a grid search over the following parameters:  (1) the kernel size of the 1-D convolutional layers $\{1, 2, 4, 8\}$; (2) the number of output channels, chosen from $\{16, 32\}$ for the first layer, $\{32, 64\}$ for the second layer, and $\{5, 10\}$ for the final layer; (3) the hidden dimensions of the last two layers of the multi-layer perceptron (MLP) $\{400, 1000\}$. The optimal configuration, determined by minimizing test error, was found to be a kernel size of $1$, output channel sizes of $\{64, 32, 5\}$, and an MLP hidden dimension of $400$. 

\subsection{Prototypical loss and SNR loss}
For the SNR loss, we conduct a grid search over batch composition and learning rate. Following the setup used in cross-entropy training, we maintain a total batch size of $512$ and explore three batch compositions: \(\{32\times16, 16\times32, 8\times64\}\) ($m$ classes \(\times\) $P$ samples per class). Additionally, we test three learning rates: $\{0.1, 0.01, 0.001\}$. Model performance is evaluated based on test error across three held-out rule splits, distinct from the $15$ used for training. The optimal configuration is $32\times16$ with a learning rate of $0.01$.

The prototypical loss~\cite{snell2017prototypical} is a distance-based loss function commonly used in few-shot learning. It generates rule prototypes, which act as representative representations for each rule. Like the SNR loss, each input batch consists of $P$ rows drawn from $m$ different relational rules. For each rule $r$, the $P$ rows are randomly split into two disjoint subsets: a support set $S_r$ and a query set $Q_r$. The prototype $z_r$ for rule $r$ is computed as the mean representation of its support set:
\begin{equation}
	z_{r}=\frac{1}{\mid S_{r}\mid}\sum_{x\in S_{r}}x
        \label{eq:Proto_z} 
\end{equation}
where $x$ denotes the relational representation of an input row. 

The prototypical loss encourages rows in the query set $Q_r$ to be closer to their corresponding rule prototype $z_r$ while pushing them away from prototypes of other rules $z_{c'} (c' \neq c)$ in the batch. It is formulated as:
\begin{equation}
	l_{\text{Prototypical}}=\sum_{r}\sum_{x\in Q_{c}}-\log\frac{\exp\left(-\parallel x-z_{c}\parallel^{2}\right)}{\sum_{c' \neq c}\exp\left(-\parallel x-z_{c'}\parallel^{2}\right)}
	\label{eq:Proto_loss} 
\end{equation}

To optimize the Prototypical loss, we conducted a grid search over:  (1) batch size \(\{32\times16, 16\times 32, 8 \times 64\}\) ($m$ classes \(\times\) $P$ samples per class); (2) learning rate {\(\{0.1, 0.01, 0.001\}\)}, and support set size {\(\{1, 5\}\)}. Model performance was assessed using test accuracy on three held-out rule splits, distinct from the 15 used for training. The optimal configuration was found to be \( 32 \) classes \(\times\) \( 16 \) samples per class, with a learning rate of \( 0.01 \) and support set size \( 5 \).

\section{Supplementary Figures}

\begin{figure}[h!]
\centering
\includegraphics[width=0.8\textwidth]{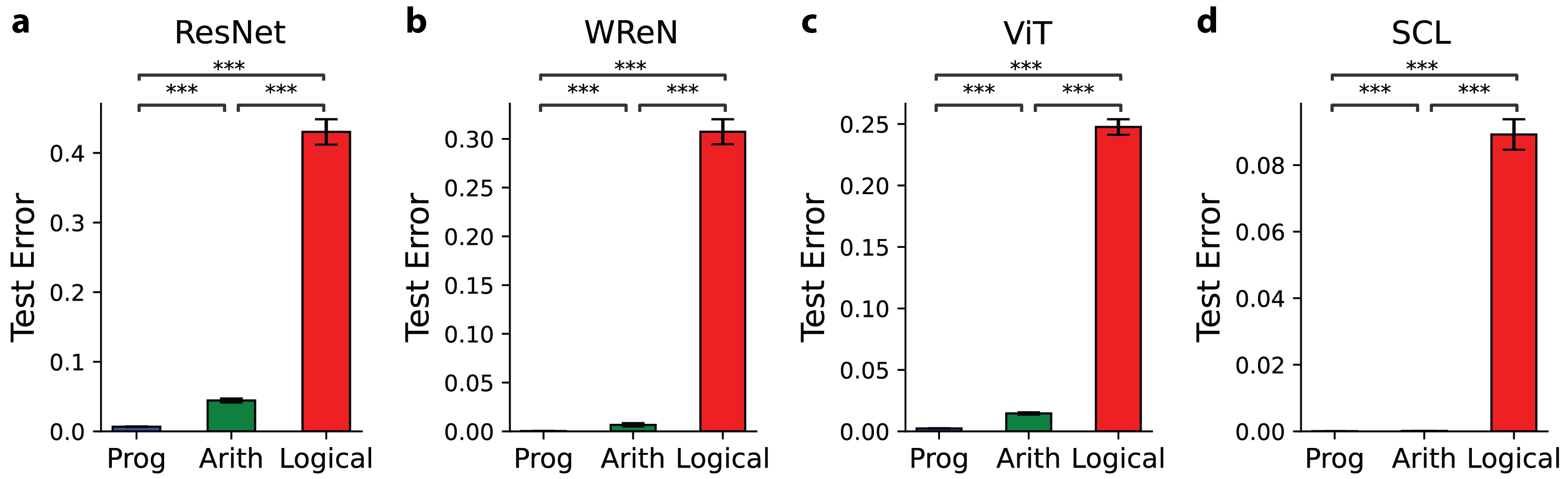}
\caption{Model Test Errors Across Abstract Relation Types.
(a-d) Test errors of four models on the $35$ training rules, evaluated across three abstract relation types: Progression (blue), Arithmetic (green), and Logical (red). The chance error is $0.97$. Each bar shows the mean classification error for rows of rules belonging to one of these relation types. Error bars represent the standard error of the mean (s.e.m.). Statistical significance was assessed using pairwise Wilcoxon tests, with asterisks indicating significance levels: *** ($p < 0.001$).}
\label{fig:SuppFig1_test_err_per_relation}
\end{figure}

\begin{figure}
\centering
\includegraphics[width=0.8\textwidth]{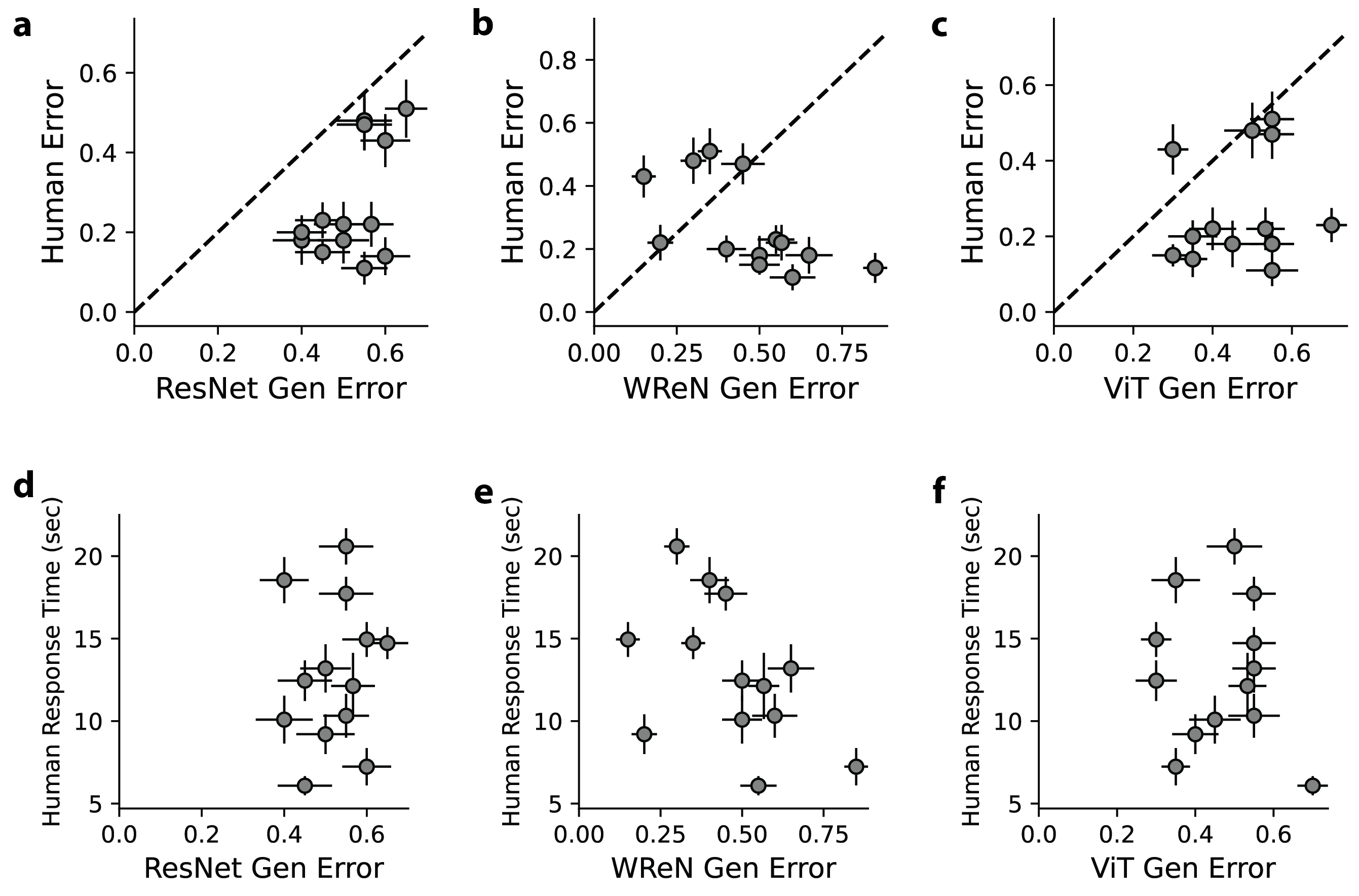}
\caption{Comparison of Model Generalization Errors with Human Performance on the \textit{SimplifiedRPM} Task.
(a-c) Comparison of the generalization errors of ResNet-50, WReN, and ViT models with human performance on the \textit{SimplifiedRPM} task across 13 selected rule pairs. Each dot represents the average model error (x-axis) and human error (y-axis) over 10 questions for a given rule pair, with error bars indicating the standard error of the mean (s.e.m.). The dashed line represents the line of identity. No significant positive correlation is observed between model and human errors (Spearman’s rank correlation: ResNet-50, $\rho = 0.34$, $p = 0.24$; WReN, $\rho = -0.67$, $p = 0.01$; ViT, $\rho = 0.27$, $p = 0.37$).
(d-f) Relationship between model generalization error (x-axis) and human response time (y-axis) for the 13 selected rule pairs in ResNet-50, WReN, and ViT models. Each dot represents the average model error and corresponding human response time for a given rule pair, with error bars indicating the s.e.m. No significant correlation is found between model error and human response time (Spearman’s rank correlation: ResNet-50, $\rho = 0.12$, $p = 0.68$; WReN, $\rho = -0.50$, $p = 0.08$; ViT, $\rho = -0.14$, $p = 0.65$).}
\label{fig:SuppFig2_human_model}
\end{figure}

\begin{figure}
\centering
\includegraphics[width=\textwidth]{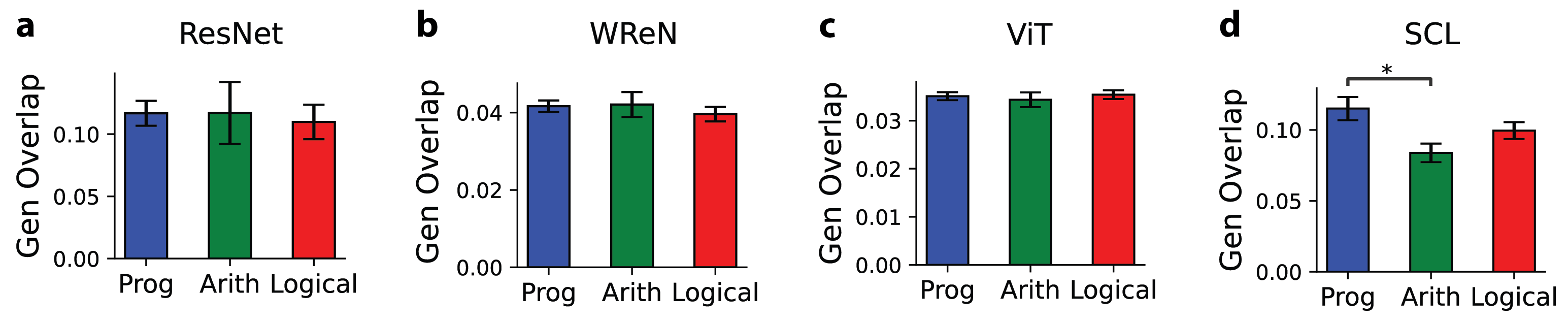}
\caption{Generalization signal-noise overlap differences across abstract relation types.
(a-d) Generalization signal-noise overlap (normalized by signal magnitude) for different abstract relation types—progression, arithmetic, and logical—across four models. Results are averaged over 15 random splits of training and held-out rules. Each bar represents the average error for trials where the correct rule of the sample row belonged to each relation type, with error bars indicating the s.e.m. Statistical comparisons were conducted using the pairwise Wilcoxon test. Asterisks indicate statistical significance: * ($p < 0.01$).}
\label{fig:SuppFig3_overlap}
\end{figure}

\begin{figure}
\centering
\includegraphics[width=0.7\textwidth]{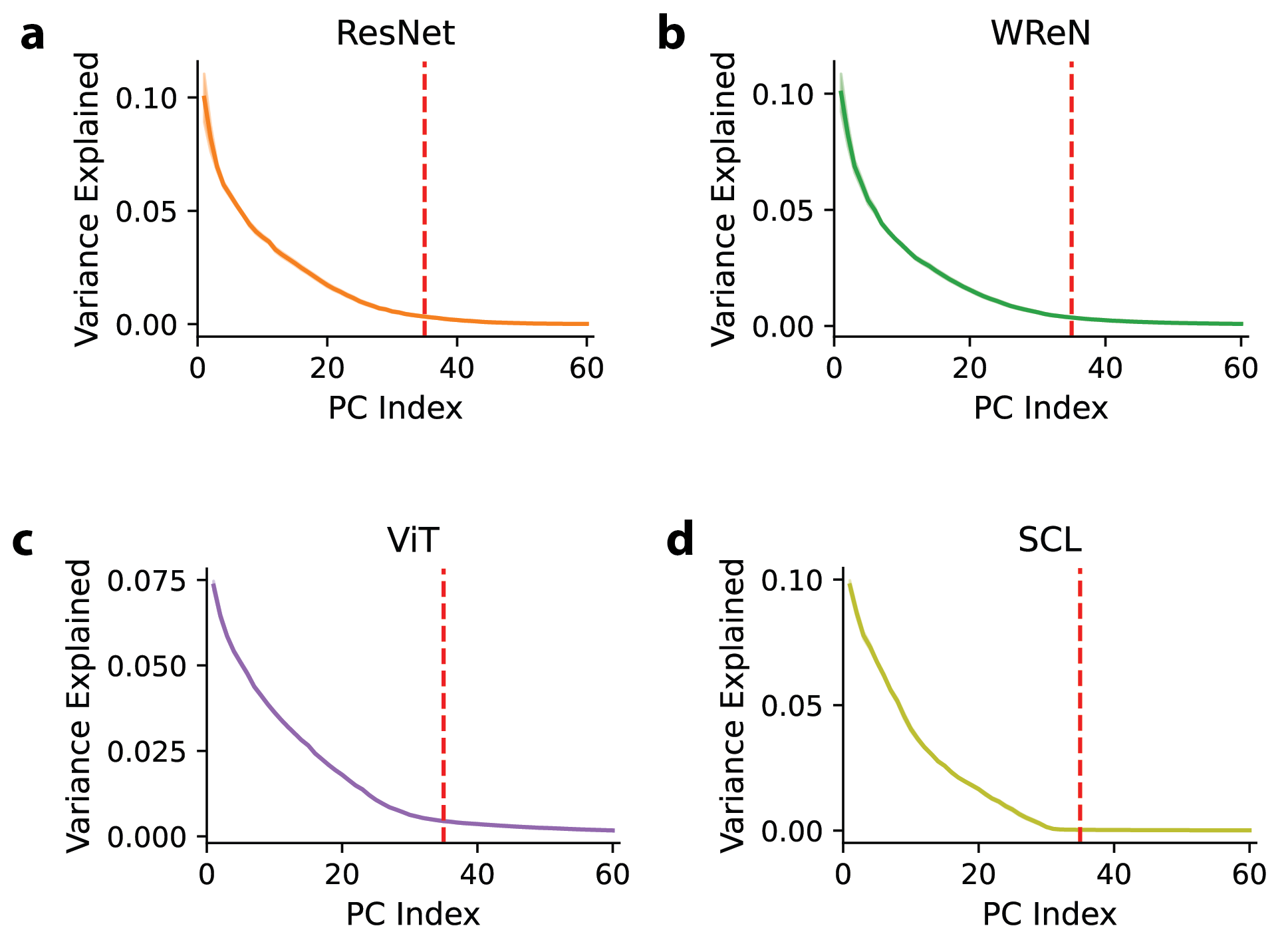}
\caption{Fraction of variance explained by the principal components (PCs) of the training rule representations.
(a-d) We aggregate the relational representations of the 1,000 test rows for each of the 35 training rules and perform principal component analysis (PCA) on these representations. For each of the four neural networks under study, we plot the fraction of variance explained by each principal component (PC), displaying only the top 60 PCs. The red line indicates the 35th PC, corresponding to the number of training rules. A noticeable drop in explained variance at this point suggests that the representations of the 35 training rules are consistently confined to a low-dimensional subspace. The lines represent the average explained variance across 15 random splits of held-out rules, with shading indicating the standard error of the mean (s.e.m.).}
\label{fig:SuppFig4_test_PC_sig}
\end{figure}

\begin{figure}
\centering
\includegraphics[width=0.8\textwidth]{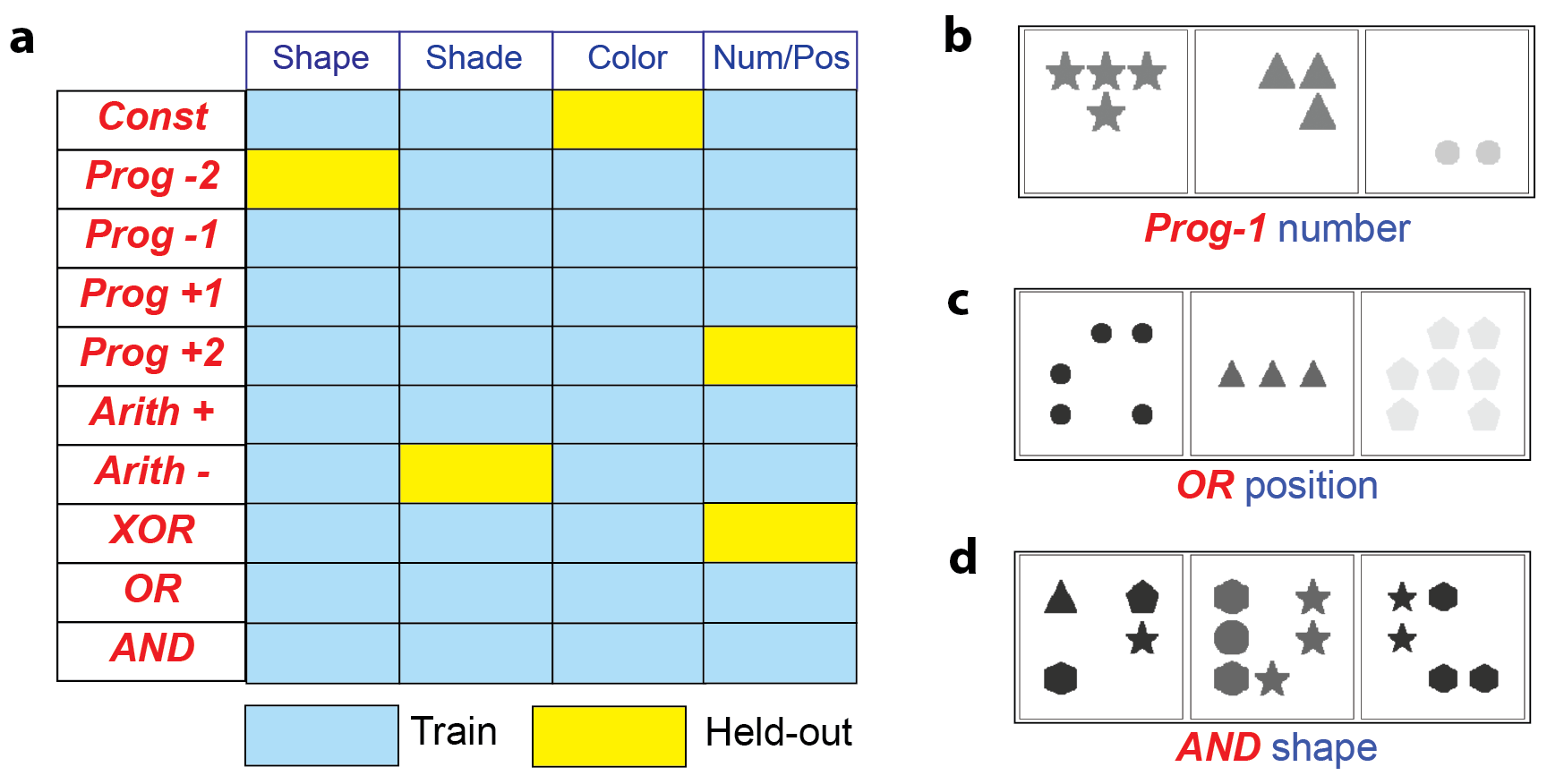}
\caption{Overview of \textit{SimplifiedRPM} dataset rules and example rows.
(a) Rules in the \textit{SimplifiedRPM} dataset. Each rule in the \textit{SimplifiedRPM} dataset is defined by an abstract relation (first column) applied to a specific object attribute, resulting in a total of 40 possible rules. During training, we create rule splits by randomly selecting five rules to be held out, ensuring that each abstract relation appears at most once in the held-out set. The table illustrates an example of this split, where yellow cells indicate the five held-out rules, and blue cells represent the remaining 35 rules used for training the models.
 (b-d) Example rows in the \textit{SimplifiedRPM} dataset. 
 (b) Example row illustrating the Progression -1 number relation, where the number of objects in each panel decreases by one across the row (4 objects in the first panel, 3 in the second, and 2 in the last). 
 (c) Example row demonstrating the OR position rule, where objects in the third panel occupy positions in either the first or second panel.
 (d) Example row illustrating the AND shape rule, where the third panel contains shapes (hexagon, star) that appear in both the first and the second panel.}
 \label{fig:SuppFig5_rule_table_example}
\end{figure}

\begin{figure}
\centering
\includegraphics[width=0.9\textwidth]{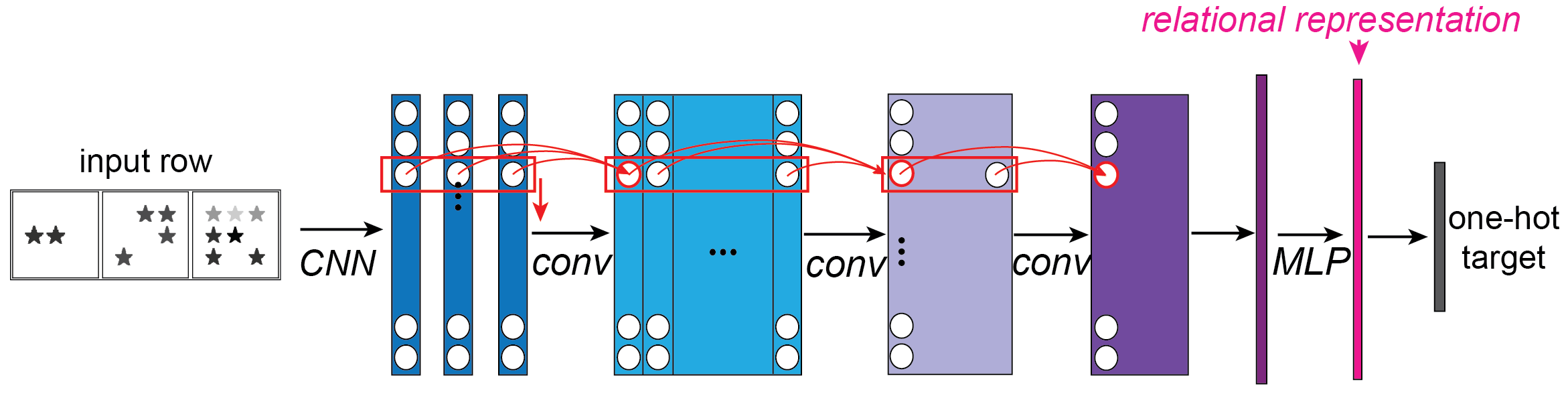}
\caption{Overview of the SCL architecture.
Each input row, consisting of three panels, is processed through three 1-D convolutional layers, followed by two fully connected multilayer perceptron (MLP) layers. The final output is of size 35, compared against the one-hot target vector corresponding to the 35 possible relational rules. The penultimate layer is the relational representation used for the \textit{SimplifiedRPM} task.}
 \label{fig:SuppFig6_SCL_architecture}
\end{figure}

\begin{figure}
\centering
\includegraphics[width=0.5\textwidth]{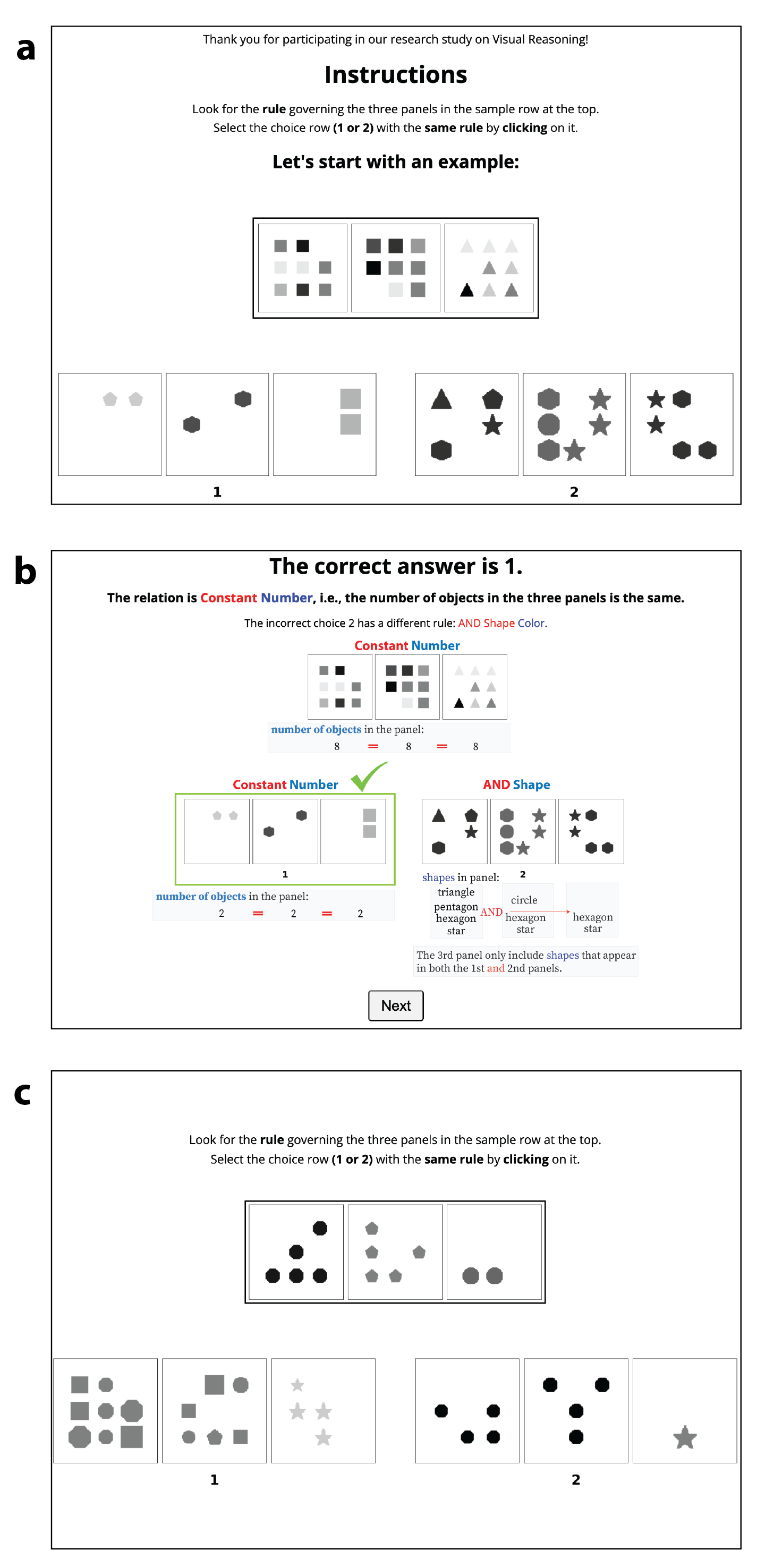}
\caption{Example trials in the human experiment.
(a) Instruction and practice trial, where participants are introduced to the task. 
(b) Explanations are provided to participants after they make a selection in the practice trial. 
(c) Example trial from the main experiment, where no feedback is given.}
 \label{fig:SuppFig7_human}
\end{figure}

\begin{figure}
\centering
\includegraphics[width=\textwidth]{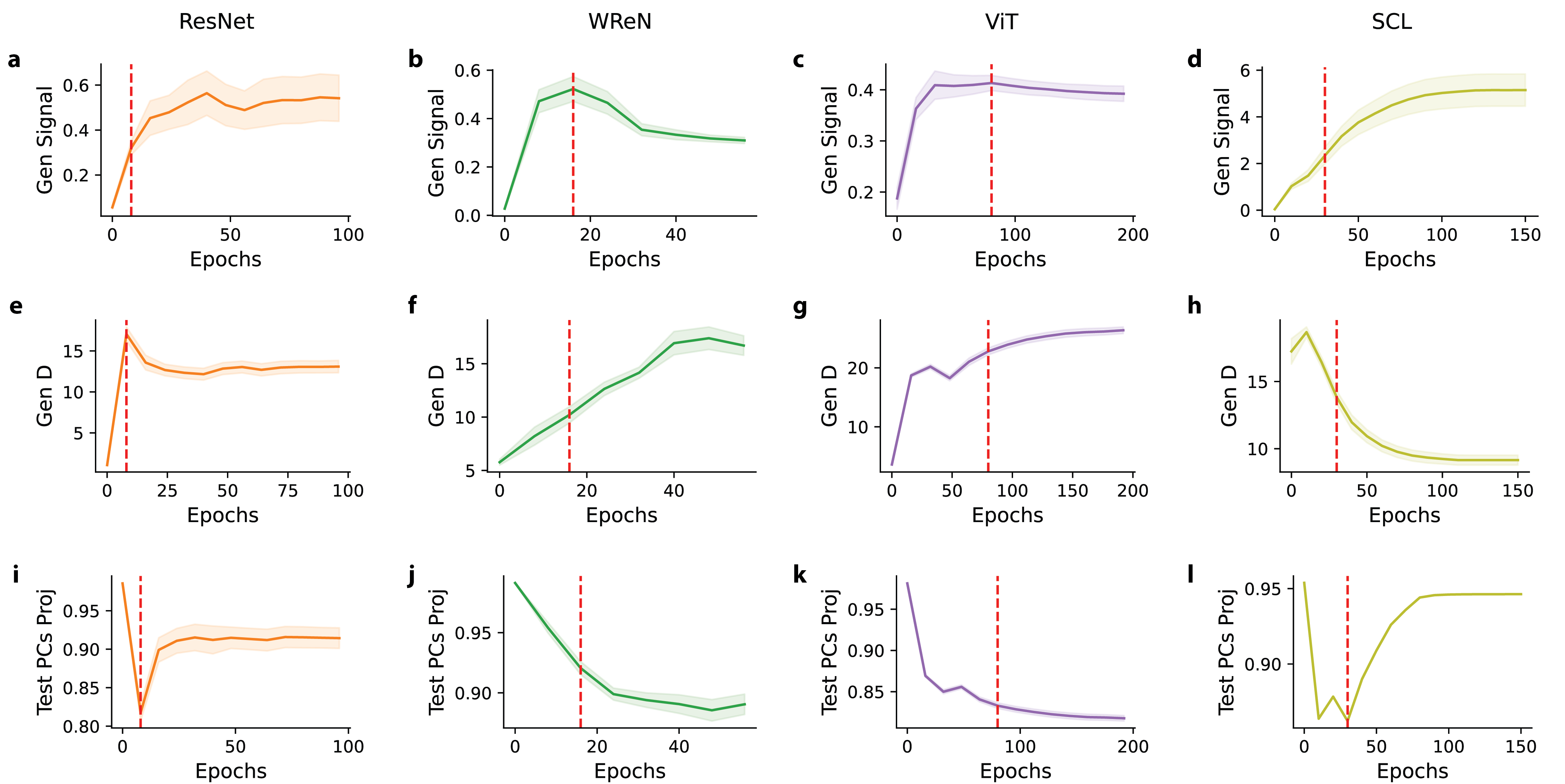}
\caption{Comparison of Model Generalization Errors with Human Performance on the \textit{SimplifiedRPM} Task.
Changes in generalization signal (a-d), generalization dimensionality D (e-h), and the variance of held-out rule manifolds explained by the top 35 principal components of training rules (i-l) over training epochs. Results are averaged across 15 different held-out rule splits, with shaded areas representing the standard error of the mean (s.e.m.) for the four models (ResNet, WReN, ViT, and SCL). The red vertical line marks the epoch at which the relational representation structure for the training rules stabilizes.}
\label{fig:SuppFig8_score_epoch}
\end{figure}

\clearpage
\section{Supplementary Tables}

\begin{table}[h!] 
	\centering
	\caption{Human Performance on 13 Rule Pairs.
	Thirteen rule pairs were tested in the human experiment. The correct rule corresponds to the rule governing both the sample row and the correct choice row, while the incorrect rule applies to the incorrect choice row. Human error represents the average error rate across 10 trials for each rule pair, with error bars indicating the standard error of the mean (s.e.m.). We recruited 25 MTurk participants (n=25), each completing a subset of 60 trials from a total of 130 (13 rule pairs $\times$ 10 trials per rule pair). The error for each trial was calculated as the mean response error across 10 participants. Response time denotes the total duration from trial onset (when the stimulus appears on the screen) to the moment the participant selects one of the two choice images. Rule pairs are ordered by average human error.}
	\label{tab:human_error} 

	\begin{tabular}{lccccr} 
		\\
		\hline
		\# & Correct Rule & Incorrect Rule & Human Error & Response Time (sec) \\
		\hline
		1  & Prog-1 Number  & AND Shape       & $0.11\pm0.04$  & $10.32\pm1.33$ \\
		2  & Prog+1 Number  & Const Number    & $0.14\pm0.05$  & $7.23\pm1.13$  \\
		3  & OR Shape       & Arith+ Number   & $0.15\pm0.03$  & $13.52\pm1.08$ \\
		4  & Prog+2 Number  & Const Number    & $0.18\pm0.06$  & $10.08\pm1.45$ \\
		5  & Arith- Number  & OR Shape        & $0.18\pm0.06$  & $13.19\pm1.46$ \\
		6  & XOR Shape      & AND Position    & $0.20\pm0.04$  & $18.57\pm1.57$ \\
		7  & Prog-2 Number  & XOR Position    & $0.22\pm0.06$  & $8.87\pm1.23$  \\
		8  & Arith+ Number  & OR Position     & $0.22\pm0.06$  & $12.13\pm2.01$ \\
		9  & Const Number   & Prog+2 Number   & $0.23\pm0.04$  & $6.08\pm0.58$  \\
		10 & AND Position   & Prog-2 Number   & $0.43\pm0.07$  & $14.95\pm1.05$ \\
		11 & AND Shape      & XOR Shape       & $0.47\pm0.06$  & $17.80\pm1.07$ \\
		12 & OR Position    & XOR Position    & $0.48\pm0.07$  & $20.62\pm1.18$ \\
		13 & XOR Position   & Arith- Number   & $0.51\pm0.07$  & $14.71\pm1.02$ \\
		\hline
	\end{tabular}
\end{table}

\begin{table} 
	\centering
	\caption{Relations in the \textit{SimplifiedRPM} Dataset. The \textit{SimplifiedRPM} dataset includes ten abstract relations organized into progression, arithmetic, and logical types. Each rule defines how attribute index values assigned to objects change across three panels. Each example consists of three lists, each representing a panel and specifying the assigned attribute values. For instance, $\left\{ [1],[1],[1]\right\}$ represents a constant relation, with objects maintaining the attribute value one across all panels. Objects can also have multiple attribute values, as seen in the \textit{AND} relation example $\left\{ [1],[0,1],[1]\right\}$, where the second panel contains objects with attributes 0 and 1.}
	\label{tab:simplifiedRPM_rules} 

        \begin{tabular}{lccc}
            \\
            \hline
            \textbf{Relation Type} & \textbf{Relation} & \textbf{Description} & \textbf{Example} \\
            \hline
            \textbf{Progression} & Const & \shortstack{Attribute values are the same \\ across all three panels.} & \{[1], [1], [1]\} \\
            \cmidrule(lr){2-4} & Prog -2  & \shortstack{Attribute values decrease \\ incrementally by 2 across the panels.} & \{[4], [2], [0]\} \\
            \cmidrule(lr){2-4} & Prog -1  & \shortstack{Attribute values decrease \\ incrementally by 1 across the panels.} & \{[2], [1], [0]\} \\
            \cmidrule(lr){2-4} & Prog +1  & \shortstack{Attribute values increase \\ incrementally by 1 across the panels.} & \{[0], [1], [2]\} \\
            \cmidrule(lr){2-4} & Prog +2  & \shortstack{Attribute values increase \\ incrementally by 2 across the panels.} & \{[0], [2], [4]\} \\
            \hline
            \textbf{Arithmetic}  & + & \shortstack{Attribute values in the 3rd panel \\ are the sum of the values \\ in the first two panels.} & \{[0], [1], [1]\}\\
            \cmidrule(lr){2-4} & - & \shortstack{Attribute values in the 3rd panel \\ are the first panel's values \\ minus the second panel's values.} & \{[1], [1], [0]\}\\
            \hline
            \textbf{Logical} & AND & \shortstack{3rd panel contains attribute values \\ present in both the first and second panels.} & \{[0], [0, 1], [0]\}\\
            \cmidrule(lr){2-4} & OR & \shortstack{3rd panel contains attribute values \\ present in either the first or second panel.} & \{[0], [0, 1], [0, 1]\}\\
            \cmidrule(lr){2-4} & XOR       & \shortstack{3rd panel contains attribute values \\ present in only one of the first two panels.} & \{[0], [0, 1], [1]\}\\
            \hline
        \end{tabular}
\end{table}

\clearpage

\end{document}